\documentclass[aps, amsfonts, amssymb, amsmath, reprint, nofootinbib, twoside, superscriptaddress]{revtex4-2}

\usepackage[english]{babel}
\usepackage[utf8]{inputenc}
\usepackage{times} 

\usepackage[colorinlistoftodos, color=green!40, prependcaption]{todonotes}
\usepackage{graphicx}
\usepackage{xfrac}
\usepackage{bm}
\usepackage{float}
\usepackage{newfloat,algcompatible}
\graphicspath{{Figures/}}
\usepackage[pdftex, pdftitle={Article}, pdfauthor={Author}]{hyperref} 
\DeclareFloatingEnvironment[
    fileext=loa,
    listname=List of Algorithms,
    name=ALGORITHM,
    placement=tbhp,
]{algorithm}

\begin{document}

\title{Bridging the reality gap in quantum devices with physics-aware machine learning}

\author{D.L.~Craig$^\dag$}
\affiliation{Department of Materials, University of Oxford, Parks Road, Oxford OX1 3PH, United Kingdom}
\author{H.~Moon$^\dag$}
\affiliation{Department of Materials, University of Oxford, Parks Road, Oxford OX1 3PH, United Kingdom}
\author{F.~Fedele}
\affiliation{Department of Materials, University of Oxford, Parks Road, Oxford OX1 3PH, United Kingdom}
\author{D.T.~Lennon}
\affiliation{Department of Materials, University of Oxford, Parks Road, Oxford OX1 3PH, United Kingdom}
\author{B.~Van Straaten}
\affiliation{Department of Materials, University of Oxford, Parks Road, Oxford OX1 3PH, United Kingdom}
\author{F.~Vigneau}
\affiliation{Department of Materials, University of Oxford, Parks Road, Oxford OX1 3PH, United Kingdom}
\author{L.C.~Camenzind}
\affiliation{Department of Physics, University of Basel, 4056 Basel, Switzerland}
\author{D.M.~Zumb\"uhl}
\affiliation{Department of Physics, University of Basel, 4056 Basel, Switzerland}
\author{G.A.D.~Briggs}
\affiliation{Department of Materials, University of Oxford, Parks Road, Oxford OX1 3PH, United Kingdom}
\author{M.A.~Osborne}
\affiliation{Department of Engineering Science, University of Oxford, Parks Road, Oxford OX1 3PJ, United Kingdom}
\author{D.~Sejdinovic}
\affiliation{Department of Statistics, University of Oxford, 24-29 St Giles, Oxford OX1 3LB, United Kingdom}
\author{N.~Ares$^*$}
\affiliation{Department of Engineering Science, University of Oxford, Parks Road, Oxford OX1 3PJ, United Kingdom}

\begin{abstract}
The discrepancies between reality and simulation impede the optimisation and scalability of solid-state quantum devices. Disorder induced by the unpredictable distribution of material defects is one of the major contributions to the reality gap. We bridge this gap using physics-aware machine learning, in particular, using an approach combining a physical model, deep learning, Gaussian random field, and Bayesian inference. This approach has enabled us to infer the disorder potential of a nanoscale electronic device from electron transport data. This inference is validated by verifying the algorithm's predictions about the gate voltage values required for a laterally-defined quantum dot device in AlGaAs/GaAs to produce current features corresponding to a double quantum dot regime.

\end{abstract}

\maketitle

\def\thefootnote{$\dag$}\footnotetext{These authors contributed equally to this work and are listed in alphabetical order.}

\def\thefootnote{$*$}\footnotetext{natalia.ares@eng.ox.ac.uk} 

\section{Introduction}

Differences between theory and experiment pervade all of science, and are one of the driving forces of human discovery. Simulations often require fewer resources than real experiments but rarely capture the full complexity of a system, limiting their practical application. Narrowing the gap between a model and the real world is key for the control of complex systems using machine learning, especially when a machine learning model is trained on a simulation before being applied to real systems \cite{tremblay2018training,peng2018sim}. The reality gap is widened further when there are quantities which are not directly observable. Such unobservable quantities may be estimated through their influence on other characteristics of the system; for example, indirect observation of black holes \cite{webster1972cygnus}, observation of the signature of Higgs boson decay \cite{aad2012observation}, or machine learning estimation of human poses from behind walls \cite{zhao2018through}. 


Solid-state quantum devices of nominally identical design will often display different characteristics. This variability hinders the scalability of otherwise promising qubit realisations, such as in the spin states of electrons confined in electrostatically-defined quantum dots \cite{petta2005coherent,borselli2011pauli,klos2018calculation}. Different devices exhibit different electron transport features for identical gate voltage values. This variability is even observed in the same device after being exposed to thermal cycling~\cite{moon2020machine}. In particular, electrostatic disorder induced by randomly located donor ions can be a significant source of variability in delta-doped semiconductor quantum dot devices \cite{stopa1996quantum,nixon1991breakdown}. Confinement potentials of individual quantum dots have been probed using in-plane magnetic fields \cite{camenzind2019spectroscopy}, but there has been no quantitative experimental study of the disorder present in these devices beyond the observation of its effects~\cite{croot2019gate}. 

To access the disorder characteristics that can only be observed indirectly through the transport of electrons, in this work we develop a physics-aware machine learning approach. We use transport measurements of an electrostatically-defined quantum dot device in an AlGaAs/GaAs heterostructure to inform and verify our approach. 

To infer the disorder potential we use a combination of transport measurements and predictions from a physical model. The physical model is an electrostatic simulation from which transport features can be estimated. Many simulations with different parameter settings are required to compare this physical model with transport measurements. To accommodate this need without extreme computation times, we develop a fast approximation of the model using deep learning. 

The transport measurements and electrostatic simulations inform the inference algorithm to produce plausible disorder potentials, i.e. posterior samples. The inference mechanism used in this paper follows the philosophy of approximate Bayesian computation~\cite{Sunnaker03, Beaumont19, Barber15, Marko19} by utilising the deep learning approximation of the electrostatic model.

A naive implementation of this inference still leads to unrealistically expensive and wasteful computation. This is because electrons are confined in a 2-dimensional electron gas (2DEG), and thus the disorder potential to be inferred is a dense 2D function. We develop a novel reparameterisation to greatly reduce the dimensionality of the inference problem, while selecting only the most informative regions of the disorder potential. This maps the non-parametric 2D disorder potential into a parametric model. This reparameterisation approximates the function in the spatial and spectral domains simultaneously using an inducing point approximation of a Gaussian process~\cite{pmlr-v5-titsias09a, BuiYT17, Bauer06}, and random Fourier features~\cite{Li19k, Avron17a, Hensman2018, Sriperumbudur2015, Choromanski18a, Recht2007}. 

To assess the performance of inference results we use the disorder potentials produced by the algorithm to predict the electron transport regime of new measurements. These predictions provide good agreement with experiment, indicating that our physics-aware method is effective. The physical model can determine the number of quantum dots at a given voltage location. Using posterior disorder samples within this model allows us to predict the voltage locations of double quantum dots and verify these predictions with the experiment. Results show that our physics-aware machine learning provides a clear advantage over an uninformed model of the disorder potential when predicting the location of double quantum dot features in gate voltage space.

\begin{figure}[ht]

    \centering
    \includegraphics[width=0.42\textwidth]{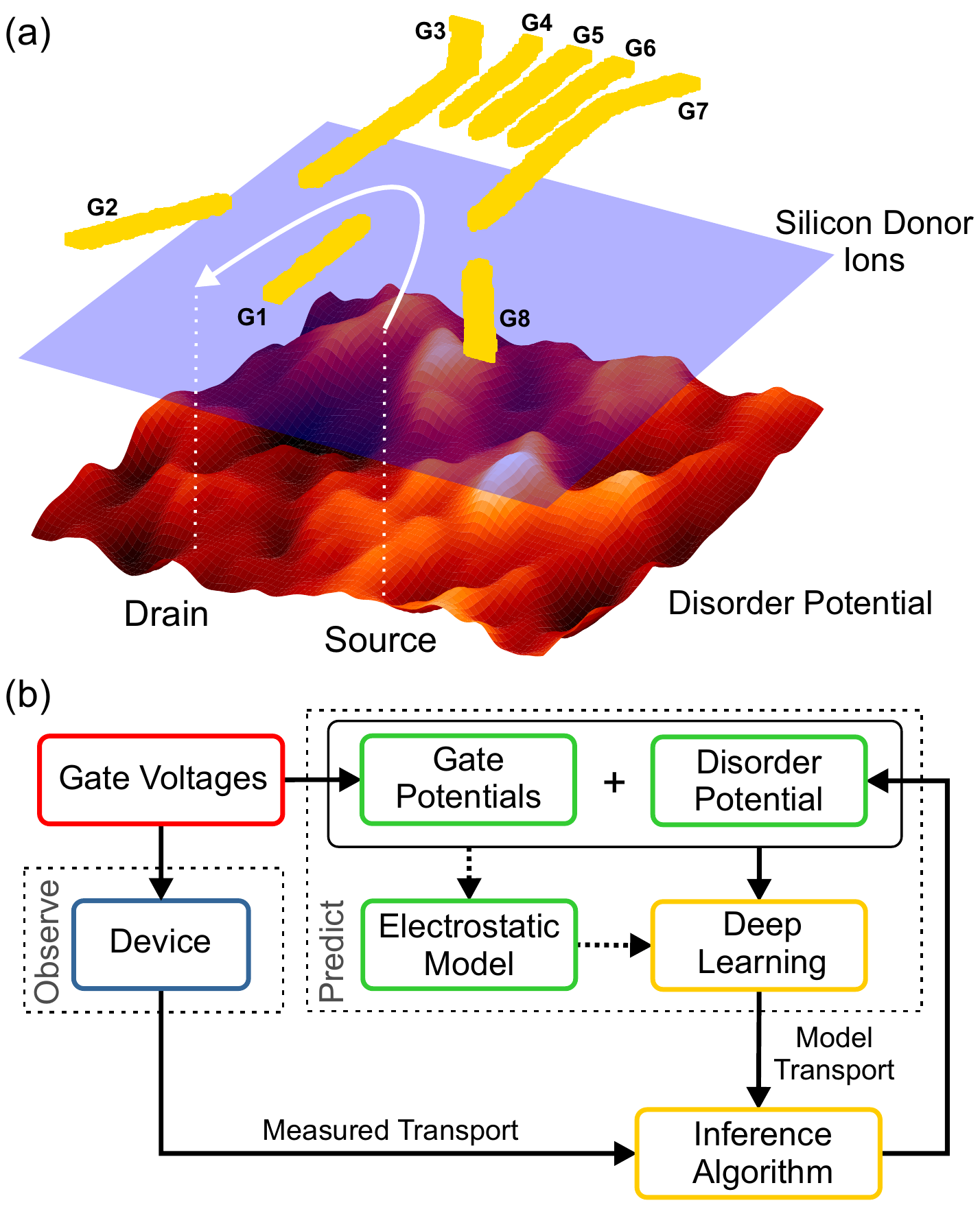}

\caption{ (a) Device geometry including the gate electrodes (labelled G1-G8), donor ion plane, and an example disorder potential experienced by confined electrons. Typical flow of current from source to drain is indicated by the white arrow. (b) Schematic of the disorder inference process. Colours indicate the following; red for experimentally controllable variables, green for quantities relevant to the electrostatic model, blue for experimental device, and yellow for machine learning methods. Dashed arrows represent the process of generating training data for the deep learning approximation and are not part of the disorder inference process. }
\label{Fig_1}
\end{figure}

\section{The Device}
\label{sec:device_exp}

A bias is applied to ohmic contacts to drive current through the device from source to drain, and applying voltages to the gates allows for the control of this current. With appropriate gate voltages, electrons may be confined to form quantum dots. Current peaks as a function of gate voltage in transport measurements are a signature of Coulomb blockade, indicating the formation of quantum dots. A random distribution of Si donor ions contributes a disordered component to the electrostatic potential experienced by electrons confined in a 2DEG. The distribution of donor ions is thought to freeze at low temperatures with rearrangement only possible significantly above device operating temperature \cite{palm1995effects}.

Our device has 8 Ti/Au gate electrodes to which DC voltages can be applied to control electron transport in a 2DEG within a GaAs/AlGaAs heterostructure \cite{camenzind2018hyperfine,moon2020machine}. The gate architecture of the device used in the experiments is depicted in Figure~\ref{Fig_1}(a), where each of the gate voltages can be set to any value between 0V and -2V. In our device gate G6 is held at 0V to avoid leakage currents. The device is operated at millikelvin temperatures.

\section{Device Model}
\label{sec:device}
\subsection{Electrostatics}
\label{subsec:elec}

As part of our physics-aware machine learning method, summarised in Figure~\ref{Fig_1}(b), we require a model of the quantum dot device. The effects of gate electrodes and donor ions on the electron density in the 2DEG are calculated self-consistently using the pinned surface model \cite{davies1995modeling,nixon1990potential}. Delta doping results in donor ions being randomly located in a plane at a constant height of 45nm above the 2DEG. With $\mathbf{r}\! =\! (x,y)$ denoting a location in the 2DEG plane, the total electrostatic potential is
\begin{equation}
\label{potential_split}
\phi_\mathrm{tot}(\mathbf{r}) = \phi_{\mathrm{g}}(\mathbf{r}) + \phi_{\mathrm{d}}(\mathbf{r}) + \phi_{\mathrm{s}}(\mathbf{r}) + \phi_{\mathrm{e}}(\mathbf{r}),
\end{equation}
where the electrostatic potential contributions are $\phi_{\mathrm{g}}$ from the gate electrodes, $\phi_{\mathrm{d}}$ from the randomly located donor ions, $\phi_{\mathrm{s}}$ from surface states, and $\phi_{\mathrm{e}}$ from the presence of electrons in the 2DEG.

\begin{figure*}[ht]

    \centering
    \includegraphics[width=0.95\textwidth]{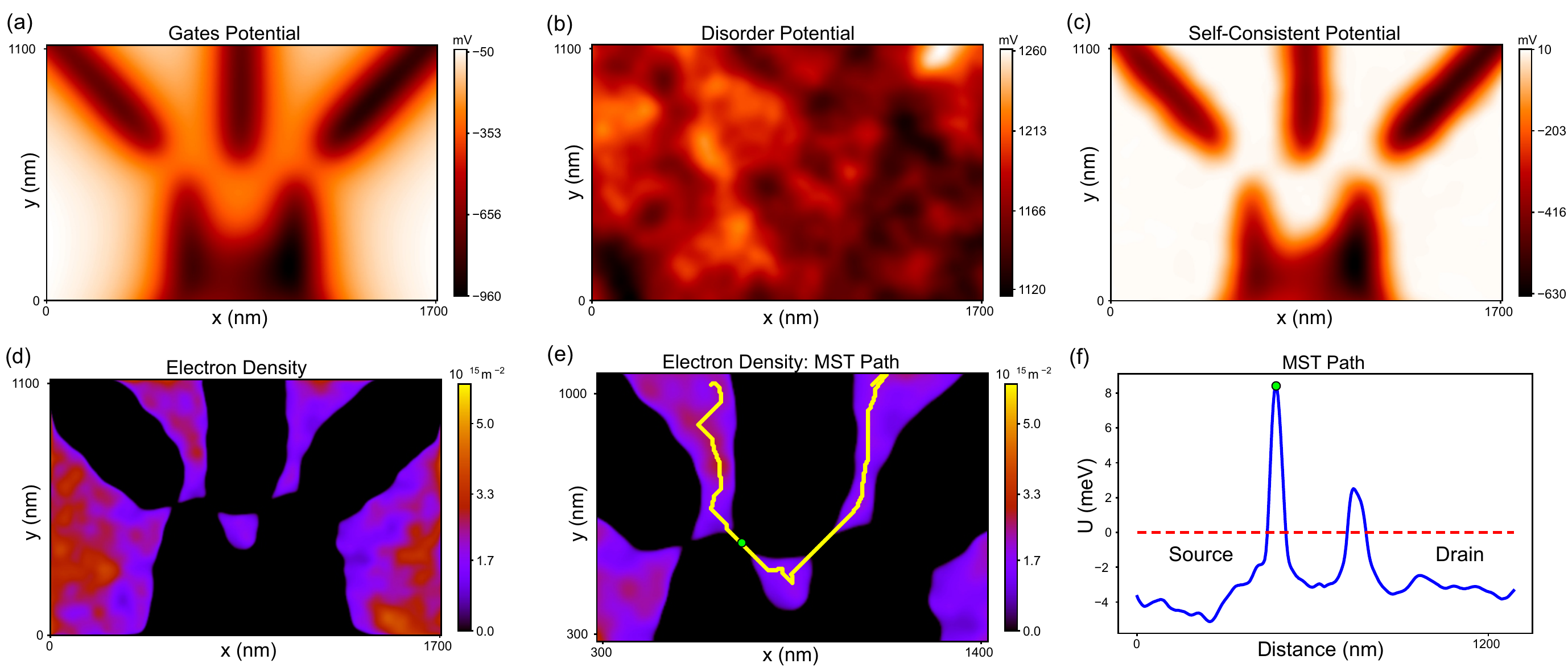}

\caption{ An example of the device model as discussed in Section \ref{sec:device}. Spatial coordinates $x$ and $y$ are used to indicate the scale of the device. (a) Electrostatic potential from the gate electrodes $\phi_{\mathrm{g}}$, (b) a disorder potential $\phi_{\mathrm{d}}$, and (c) self-consistent potential $\phi_{\mathrm{tot}}$ given the potentials in (a) and (b). (d) The electron density in the 2DEG given the potential in (c), which shows a single dot. (e) Example of MST path from source to drain in 2D (yellow line) with the location of $U^*$ is marked by a green circle. (f) The potential energy, $U$, corresponding to the MST path in (e) with $U^*$ marked by a green circle. The horizontal axis indicates the total distance moved in 2D space. The channel is closed since the value of $U^*$ is above the Fermi level indicated by the red dashed line. }
\label{Fig_2}
\end{figure*}

The potential $\phi_{\mathrm{g}}$ results from the combined effect of the potential from each gate electrode weighted by the applied voltages. We find that this model underestimates the magnitude of $\phi_{\mathrm{g}}$, so we use experimental data to fit an appropriate scale factor for each thermal cycle as discussed in Appendix~\ref{algo_appendix}. The surface potential is determined by the Schottky barrier with the gates, as discussed by Ref.~\cite{davies1988electronic}. Following this work, we set the surface potential to a constant value of $\phi_{\mathrm{s}}=-800\mathrm{mV}$. The potential in the 2DEG from a donor at location $\mathbf{r}_{k}$ in the donor plane is $\phi_{\mathrm{d}}(\mathbf{r},\mathbf{r}_{k})$. The random potential from all donor ions is then $\phi_{\mathrm{d}}(\mathbf{r}) = \sum_{k} \phi_{\mathrm{d}}(\mathbf{r},\mathbf{r}_{k})$, summing over the location of each donor. Examples of $\phi_{\mathrm{g}}$ and $\phi_{\mathrm{d}}$ are shown in Figure~\ref{Fig_2}(a) and (b) respectively.

Calculated using the Thomas-Fermi approximation in 2D, the electron density contributes to $\phi_\mathrm{tot}$ while also depending on $\phi_\mathrm{tot}$. A self-consistent solution for $\phi_\mathrm{tot}$ is computed using an iterative under-relaxation process, with an example shown in Figure~\ref{Fig_2}(c). 

\subsection{Modelling the Transport Regime}
\label{subsec:current_and_dots}

To model the transport regime of the device we consider the transport path of an electron from source to drain. If any point on the transport path has a fully depleted electron density, we say the classical channel for transport is closed (i.e. current does not flow freely). When the channel is closed, the device can be in the quantum dot regime with transport features from quantum tunneling events, or pinch-off where no current flows at all. When scanning a random combination of all gate voltages we can approximate the device as an open or closed channel. 

A semi-classical electron trajectory between source and drain is calculated by formulating $\phi_\mathrm{tot}$ as a graph, where each pixel is a node with nearest neighbour edges weighted by the mean of connected node values. The minimum spanning tree (MST) \cite{cormen2009introduction} of the graph is calculated and the unique path from source to drain is determined as shown in Figure~\ref{Fig_2}(e). With the electrostatic potential energy defined as $U(\mathbf{r})\! =\! -e\phi_\mathrm{tot}(\mathbf{r})$, the path through the MST has the minimum possible maximum value of $U$. The location of this point will be called the minimax point, $\mathbf{r}_*\! =\! (x_*,y_*)$, with $U^*\! \equiv\! U(\mathbf{r}_*)$. If $U^*$ is greater than or equal to the Fermi energy $\mu_F$, the model transport channel is considered closed. 

The electron trajectory approximated by the MST path can also be used to determine the number of quantum dots formed by a given $\phi_{\mathrm{tot}}$. The number of dots defined in the device can be determined using regions of the 1D MST path where $U(\mathbf{r}_*)\! <\! \mu_F$ which are delimited by barriers with $U(\mathbf{r}_*)\! \geq\! \mu_F$. An example of the electron density and path corresponding to a single dot in our model is shown in Figure~\ref{Fig_2}(d-f). Since dots in our device are 2-dimensional objects in the plane of the 2DEG, the 1-dimensional MST path from source to drain is not sufficient to fully determine the number of dots. Additional paths through the MST are calculated to ensure dot labels are robust to all possible configurations of the electron density. Transport features corresponding to quantum dots can only be observed near the closed channel boundary due to tunnel barriers typically suppressing current far beyond this boundary. The dots identified using our model are not affected by this limitation.

\subsection{Deep Learning Approximation}
\label{subsec:deep_learning}

For disorder inference we require fast prediction of the transport regime, determined by $U^*$ in our model, given gate voltages and a disorder potential. The self-consistent electrostatic model and MST path require up to 10 seconds to calculate $U^*$ in serial computation. This computation time is impractical for the large batches of $U^*$ required for the inference algorithm. Deep learning methods, and their ease of implementation on GPU hardware, allow for a significant acceleration \cite{ryczko2019deep,PhysRevA.96.042113}.

A convolutional neural network (CNN) is trained to calculate $\phi_{\mathrm{tot}}(\mathbf{r}_*)$. The architecture of a CNN is particularly suited to data in 2D grids such as the potentials in our electrostatic model. Each input is a 2D potential $\phi_{\mathrm{in}} = \phi_{\mathrm{g}} + \phi_{\mathrm{d}} + \phi_{\mathrm{s}}$ where $\phi_{\mathrm{g}}$ and $\phi_{\mathrm{d}}$ are randomly generated and $\phi_{\mathrm{s}}$ remains constant. The output training data consists of the self-consistent potential $\phi_{\mathrm{tot}}$ and $\phi_{\mathrm{tot}}(\mathbf{r}_*)$ corresponding to each input, with $U^*\! =\! -\phi_{\mathrm{tot}}(\mathbf{r}_*)$ in units of electron volts. The complete mapping, expressed as $\phi_{\mathrm{in}} \to \phi_{\mathrm{tot}} \to \phi_{\mathrm{tot}}(\mathbf{r}_*)$, is approximated by the CNN $\mathcal{F}_U$. The resolution of $\phi_{\mathrm{in}}$ and $\phi_{\mathrm{tot}}$ is reduced from that used in the electrostatic model to improve the performance of $\mathcal{F}_U$. A series of resolution preserving convolutions in a residual neural network (ResNet) architecture \cite{he2016deep} learn the non-linear transformation $\phi_{\mathrm{in}} \to \phi_{\mathrm{tot}}$ and further layers learn the mapping $\phi_{\mathrm{tot}} \to \phi_{\mathrm{tot}}(\mathbf{r}_*)$. Test results achieve a mean absolute error (MAE) of 1.27meV in $U^*$ estimations, with a 1.2\% error when classifying the transport regime using $U^*$. Batching inputs and using a GPU (GTX 1080 Ti) gives a computation time of 0.6ms for a single $U^*$ using $\mathcal{F}_U$, a speed up of order $10^4$ over the electrostatic model and path finding algorithm. This evaluation of $U^*$ is also significantly faster than measurement of current. The parallel computation of CNN outputs surpass any acceleration which could be achieved by optimising the exact computation of methods discussed in \ref{sec:device}, which cannot be parallelised.

To make predictions of voltage locations with a given number of quantum dots using disorder inference results, a fast method for counting dots is required. A second CNN is trained to approximate the number of quantum dots at a given set of gate voltages. The input is $\phi_{\mathrm{in}}$ as used for $\mathcal{F}_U$, with the output being the number of dots, $N_{\mathrm{dot}}\in\{0,1,2,3\}$. The network learns the mapping $\mathcal{F}_D : \phi_{\mathrm{in}} \to P(N_{\mathrm{dot}})$, where $P(N_{\mathrm{dot}})$ is the probability for a given $N_{\mathrm{dot}}$ and classification is determined by the maximum $P(N_{\mathrm{dot}})$. Due to the sparsity of dots in gate voltage space, the training set used for $\mathcal{F}_U$ is such that the classifier cannot accurately determine $N_{\mathrm{dot}}$, but only the presence or absence of dots. We thus use an intermediate classifier which produces a new training set that ideally only includes gate voltages for which $N_{\mathrm{dot}}>0$. A mixture of selected (dot-abundant) data and the original (dot-sparse) data is used to train $\mathcal{F}_D$. When determining the maximum number of dots in the direction of a given vector of gates voltages, $\mathcal{F}_D$ achieves $95.8\%$ classification accuracy. Using a GPU with batched inputs, the computation time for a single classification with $\mathcal{F}_D$ is 0.6ms. Further details of networks $\mathcal{F}_U$ and $\mathcal{F}_D$ can be found in the supplemental material. 

\section{Inference Algorithm}
\label{sec:inference_algorithm}

\subsection{Disorder Potential Reparameterisation}
\label{subsec:reparameterisation}

The disorder potential used in the electrostatic model is a dense 2D grid covering the entire 2DEG plane, as displayed in Figure~\ref{Fig_2}(b). A dense grid is unnecessary for inference since $\phi_\mathrm{d}$ is continuous and values can be interpolated from a sparse grid. Using a dense grid would be in unfeasible even with the reduced resolution CNN inputs.

\begin{figure}[ht]

    \centering
    \includegraphics[width=0.40\textwidth]{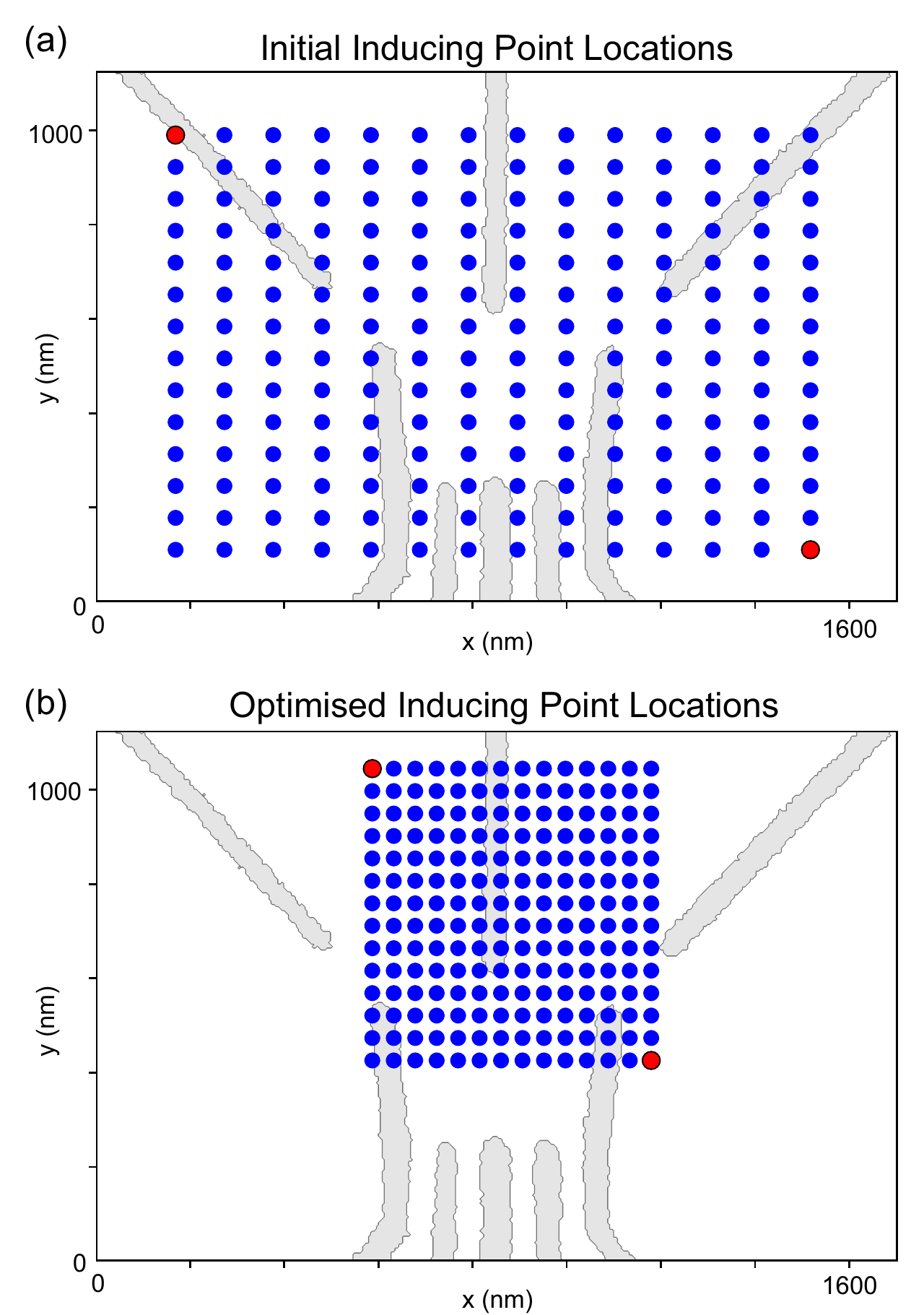}

\caption{Gate architecture overlaid with inducing point locations $X$, indicated by blue dots (a)  before optimisation and (b) after optimisation. Red circles indicate the two corners defining the rectangular array of inducing points. The location of these corners are optimised to ensure that the disorder potential values at the inducing points determines whether the transport channel is open or closed. Hence the optimised inducing points are located closer to the source and drain reservoirs.}
\label{Fig_3}
\end{figure}

We propose a novel reparameterisation algorithm, with the objective to find a set of $n_Z$ locations $Z=\lbrace \mathbf{r}^Z_k | k=1,\ldots,n_Z \rbrace$, where the disorder potential values on those locations sufficiently determine the transport regime. Following the literature of Gaussian Process regression~\cite{pmlr-v5-titsias09a}, $Z$ defines a set of inducing points. For the experiments in this paper, $Z$ is parameterised as a $14\times14$ uniform grid defined by two corner points, with the initial grid shown in Figure~\ref{Fig_3}(a). 

Our reparameterisation requires the locations of the inducing points $Z$ as well as the values of the disorder potential on these points, represented by a vector $\boldsymbol{\alpha}=[\phi_\mathrm{d}(\mathbf{r}^Z_1), \ldots , \phi_\mathrm{d}(\mathbf{r}^Z_{n_Z})]$. The full dense grid of $\phi_\mathrm{d}$ cannot be exactly recovered from the values on $Z$, because the inducing points are too sparse and random disorder potential variations between the points could influence the transport regime. This variability is encoded in the vector $\boldsymbol{\beta}=[\epsilon_1,\ldots,\epsilon_{2q}]$ containing amplitudes of random Fourier features~\cite{Recht2007, Choromanski18a, Sriperumbudur2015, Hensman2018, Avron17a, Li19k}, where $q$ is the number of frequencies considered. The parameters contained in $\boldsymbol{\beta}$ are thus dependent on $Z$.

With optimal inducing points $Z_\mathrm{opt}$, the disorder potential values contained in $\boldsymbol{\alpha}$ sufficiently determine the transport regime, while the contribution of random Fourier features from $\boldsymbol{\beta}$ is marginal. A numerical optimiser is used to find $Z_\mathrm{opt}$, where the optimisation objective is to minimise the effect of $\boldsymbol{\beta}$ on transport regime predictions made by $\mathcal{F}_U$. During optimisation, the disorder potential contributing to the input of $\mathcal{F}_U$ is approximately reconstructed from $Z$, $\boldsymbol{\alpha}$, and $\boldsymbol{\beta}$ using a deterministic function $f$ (see Appendix~\ref{sec:reparam}),
\begin{equation}
\label{phi_hat_eqn}
    \phi_\mathrm{d} \approx \hat{\phi} = f(Z,\boldsymbol{\alpha},\boldsymbol{\beta}).
\end{equation}
The optimisation of $Z$ can be performed on simulated data, and the optimised inducing points used by our inference algorithm are shown in Figure~\ref{Fig_3}(b). We observe that the inducing points are located where the transport channel is more likely to be depleted, so that the disorder potential on $Z_\mathrm{opt}$ can determine the transport regime of the device. Detailed formulation and implementation of the inducing point optimisation algorithm can be found in Appendix~\ref{algo_appendix}.

\subsection{Bayesian inference}
\label{subsec:bayesian_infernece}
To reconstruct the disorder potential, in addition to determining $Z_\mathrm{opt}$, we must infer suitable values of $\boldsymbol{\alpha}$ and $\boldsymbol{\beta}$. To do this, the inference algorithm requires measurements of current in gate voltage space. We generate random directions in the 7-dimensional gate voltage space, each defined by a unit vector $\mathbf{u}_j$ normalised such that $\max_i{\vert u_j^i \vert}=1$ and $u_j^i\leq0$ for $u_j^i\in\mathbf{u}_j$. A specific voltage location is defined as $\mathbf{v} = \mathrm{R}\mathbf{u}_j$, where $R$ is the voltage distance along $\mathbf{u}_j$. In particular, the inference algorithm requires information about the location of the boundary between open and closed channel transport. To obtain this information, stored in a dataset $D$, a current trace is conducted along a given $\mathbf{u}_j$ from the origin at $R\! =\! 0\mathrm{mV}$ to the device voltage limit at $R\! =\! 2000\mathrm{mV}$. Each current trace contributes 2 entries in $D$; the voltages immediately before and after current drops to half the open channel current, paired with $y\! =\! 1$ and $y\! =\! 0$ respectively. The resulting dataset can be defined by $D=\lbrace(\mathbf{v}_i, y_i) | i=1,\ldots,2n_\mathrm{u} \rbrace$, where $n_\mathrm{u}$ is the number of unit vectors considered.  We use $n_\mathrm{u}\! =\! 200$ in this paper, which is well below the typical data requirements of deep learning methods used to predict features of quantum devices \cite{euler2020deep,flurin2020using}.

To infer $\boldsymbol{\alpha}$ and $\boldsymbol{\beta}$ using $D$ we define a prior distribution $p(\boldsymbol{\alpha}, \boldsymbol{\beta})$ and a likelihood of data $p(D|\boldsymbol{\alpha}, \boldsymbol{\beta})$. The posterior distribution then follows the Bayes rule, $p(\boldsymbol{\alpha}, \boldsymbol{\beta} | D) \propto p(D|\boldsymbol{\alpha}, \boldsymbol{\beta}) p(\boldsymbol{\alpha}, \boldsymbol{\beta})$. In our formulation, $p(\boldsymbol{\alpha}, \boldsymbol{\beta})$ follows the multivariate normal distribution having the zero mean vector and diagonal covariance matrix. The likelihood function utilises the estimated $U^*$ from the CNN $\mathcal{F}_U$ for each data point $(\mathbf{v}_i, y_i)$ by calculating $\phi_g$ from $\mathbf{v}$, and approximating $\phi_\mathrm{d}$ from $\boldsymbol{\alpha}$ and $\boldsymbol{\beta}$.

A set of $n_\mathrm{s}$ posterior samples $\lbrace(\boldsymbol{\alpha}_i, \boldsymbol{\beta}_i) | i=1,\ldots,n_\mathrm{s}\rbrace$ can be drawn from Markov-chain Monte Carlo (MCMC) methods. Using (\ref{phi_hat_eqn}), the posterior samples of $\boldsymbol{\alpha}$ and $\boldsymbol{\beta}$ generate a set of 2D disorder potentials $S_{\hat{\phi}}\! =\! \{\hat{\phi}_i | i=1,\ldots,n_\mathrm{s}\}$, which can be used for CNN inputs. The CNN computation is differentiable, unlike the electrostatic model and path finding algorithm, allowing us to use Hamiltonian Monte Carlo (HMC) \cite{neal2011mcmc} with TensorFlow Probability \cite{dillon2017tensorflow}. 

\begin{figure}[ht]

    \centering
    \includegraphics[width=0.35\textwidth]{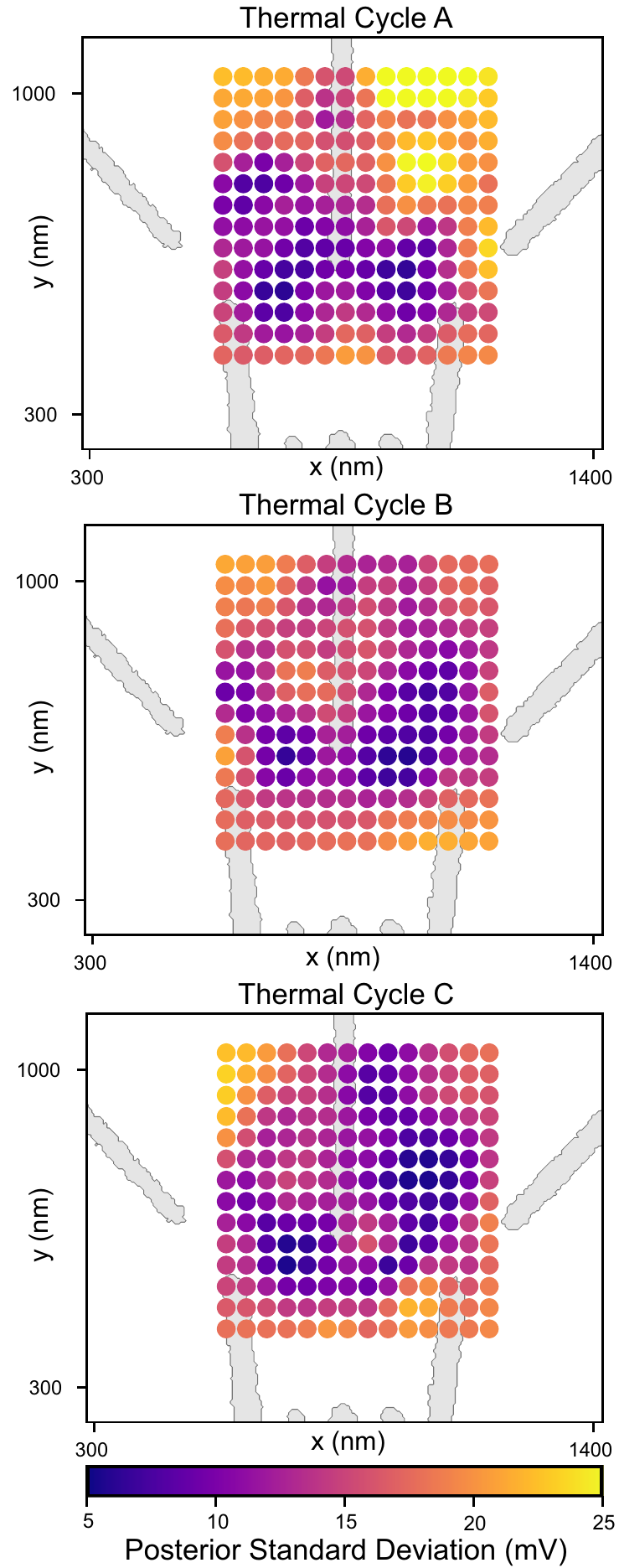}

\caption{Disorder inference results using experimental data for three thermal cycles (A,B,C) of the same device. The inducing point locations, $Z_\mathrm{opt}$, are indicated by circles with the gate structure in the background. The colour of each inducing point represents the standard deviation over posterior samples of the disorder potential value at that point.}
\label{Fig_4}
\end{figure}

\section{Results}

\subsection{Transport Channel Prediction}
\label{subsec:current_prediction}

From the Bayesian inference process we obtain a set of posterior samples of the disorder potential $S_{\hat{\phi}}$. The standard deviation of posterior inducing point values used to generate $S_{\hat{\phi}}$ is shown in Figure~\ref{Fig_4} for three thermal cycles of the same device. A low posterior standard deviation on an inducing point means the inference algorithm has learned more about the disorder potential at that location. In each case the posterior standard deviation is lowest in regions surrounding gate G1 (the `nose'). This reflects the possible locations of $U^*$ existing most frequently in these locations, due to the primary role of G1 in depleting the transport path from source to drain.

\begin{figure*}[ht]

    \centering
    \includegraphics[width=0.9\textwidth]{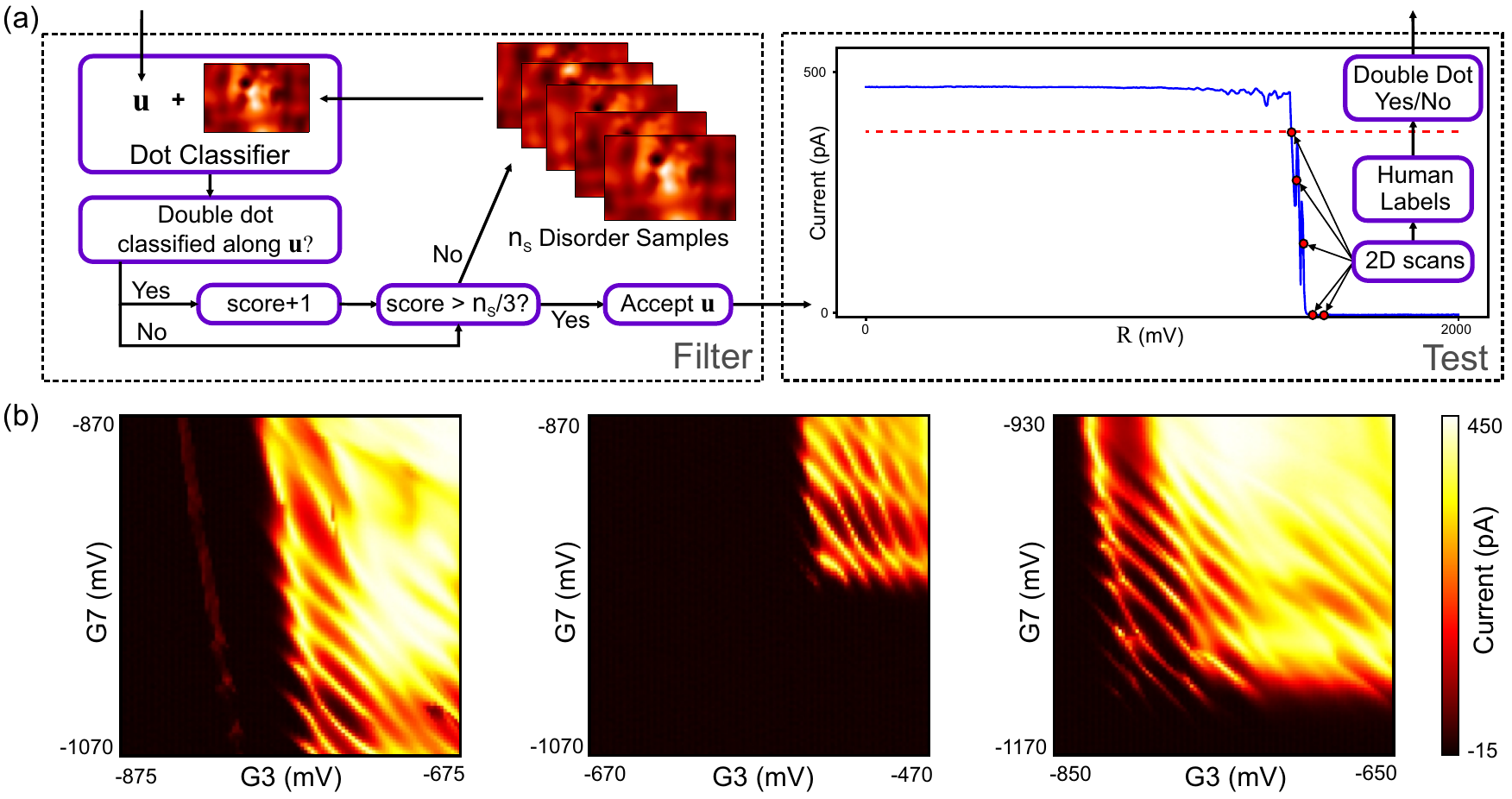}

\caption{(a) Predicting double dot locations. A unit vector $\mathbf{u}$ is passed to the filter which determines whether the vector is considered for the test device (which can be a real or simulated device). The filter uses the dot classifier $\mathcal{F}_D$ to scan along $\mathbf{v}=\mathrm{R}\mathbf{u}$ for each of the $n_\mathrm{s}$ disorder samples and the score is increased for each disorder sample which produces a double dot in the scan. Posterior disorder samples from the inference algorithm are shown. Vectors with a score greater than $\sfrac{n_\mathrm{s}}{3}$ are accepted to be tested. An example current trace (solid blue line) of a voltage vector from the origin to the limit of device operation is shown. The dashed red line indicates the $80\%$ threshold used to begin 2D current scans over gates G3 and G7. The 2D scans are taken at intervals along the original current trace (indicated by red circles). The resulting 2D current scans are passed to multiple human experts to label the presence of double quantum dots.  (b) Example current scans over G3 and G7 which scored highly for double dots when labelled by 6 human experts for 3 different unit vectors. Each scan is a 200mV$\times$200mV window with the voltages associated with the direction of $\mathbf{u}$ at the centre.}
\label{Fig_5}
\end{figure*}

After performing inference of the disorder potential, the set of posterior samples is used in the electrostatic model approximated by $\mathcal{F}_U$. We verify the posterior prediction of the distance $R_\text{C}$ required to close the transport channel in a simulated and experimental device for a set of unit vectors, $\{\mathbf{u}_i | i=1,\ldots,n_\mathrm{v}\}$. We set $R_\text{C}$ to be the point at which the current drops below $50\%$ of the open channel current. For a given $\mathbf{u}\! \in\! \{\mathbf{u}_i | i=1,\ldots,n_\mathrm{v}\}$, we calculate the mean value of $R_\text{C}$ predicted using each posterior sample in $S_{\hat{\phi}}$. To probe the generality of inference results, we evaluate predictions using the measurements which inform our inference (training data) and measurements which the inference algorithm does not encounter (test data). The training and test datasets use $n_\mathrm{v}\! =\! n_\mathrm{u}\! =\! 200$ and $n_\mathrm{v}\! =\! 300$ unit vectors respectively.

For a simulated device in which the true disorder can be chosen but is hidden from the algorithm, we compare the performance of random and posterior disorder potentials when predicting the value of $R_\mathrm{C}$ over 5 independent iterations of the inference algorithm. Random disorder potentials, generated using the electrostatic model with randomly located donor ions, predict $R_\mathrm{C}$ with a mean absolute percentage error (MAPE) of $7.0\%$ across training and test data. In contrast, posterior samples predict the value of $R_\mathrm{C}$ with a MAPE of $0.3\%$ on training data, and $0.5\%$ on test data. These results show that the inference algorithm is successful in finding disorder potentials which effectively describe features of a simulated device.

We then verify the posterior prediction of $R_\text{C}$ in a real device.  Thermal cycling the device 4 times, we run a total of 5 iterations of the inference algorithm. The value of $R_\text{C}$ is predicted with a MAPE of $1.5\%$ for training data and $2.0\%$ for test data. The MAE of $R_\text{C}$ predictions is $24.3\textrm{mV}$ and $31.9\textrm{mV}$ for training and test data respectively. Random disorder potentials predict $R_\mathrm{C}$ with a mean absolute percentage error (MAPE) of $7.5\%$ across training and test data. Compared to the simulated device, the reduced performance of the inference algorithm can be attributed to differences between the model and the experiment. The inference results remain effective in predicting the gate voltages which close the transport channel.

\subsection{Double Dot Prediction}
\label{subsec:dot_prediction}

Having demonstrated the success of the inference algorithm in determining the values of $R_\text{C}$, we use the posterior disorder samples to predict transport features beyond the training domain of the inducing point optimisation and inference algorithm. We specifically consider features corresponding to the double quantum dot regime. We implement a method requiring minimal knowledge of the transport characteristics of a particular device. Three pieces of information are required, i) quantum dots form near the closed channel boundary, ii) gates G3 and G7 in our device couple most strongly to dot energy levels, and iii) double quantum dots form features with periodicity in two gate voltage directions in transport measurements. 

The method of finding double quantum dots using posterior disorder samples is summarised in Figure~\ref{Fig_5}(a). Random unit vectors $\mathbf{u}$, are generated and scanned from $R=0\mathrm{mV}$ to $R=2000\mathrm{mV}$ in the simulated device. A randomly chosen unit vector is unlikely to lead to double dot transport features given the sparsity of double dots in voltage space. Based on predictions made by $\mathcal{F}_D$ for each posterior disorder, we select candidate voltage vectors. If $\mathcal{F}_D$ detects a double dot along a vector for a given posterior disorder, the vector's score is increased by one. Vectors with a score greater than a selected threshold (taken to be $\sfrac{n_\mathrm{s}}{3}$) are accepted to be investigated further in a test device.

Accepted vectors only indicate a direction in gate voltage space in which double quantum dot features could be observed in a test device. As these features are expected to be found near the closed channel boundary in transport measurements, we investigate multiple voltage locations near this boundary along each accepted vector. To investigate each accepted $\mathbf{u}$, an automated protocol performs a current trace along $\mathbf{u}$ from the origin to the device voltage limit. The gate voltages are then set to the boundary between open and closed channel regimes along $\mathbf{u}$, identified by a drop of 20\% from open channel current. To allow for the identification of double quantum dot transport features, gates G3 and G7 are scanned in a 200mV$\times$200mV window centred at this gate voltage location. Such 2D scans are subsequently performed at intervals of 13.3mV in $R$ along the direction of $\mathbf{u}$ until the maximum current value in a 2D scan drops below 100pA. The 2D scans are labelled by 6 human experts to determine the presence of double quantum dot features along $\mathbf{u}$. Multiple human experts are required as double quantum dot features are often subjective to human labellers and difficult to identify computationally \cite{moon2020machine}. Further details of the vector filtering and labelling of 2D scans can be found in the supplemental material.

Similar to our $R_\text{C}$ predictions, we first test the predictive power of posterior disorders in a simulated device in which the true disorder potential is known. By selecting random unit vectors we find a mean double dot occurrence rate of 0.83\% using several random disorder potentials, generated using the electrostatic model with randomly located donor ions. We thus perform disorder inference followed by vector filtering. We do not scan gates G3 and G7 as in the real device since $\mathcal{F}_D$ can determine the number of dots at a point in voltage space.  After performing vector filtering, double dots are correctly identified in 28\% of instances using random disorders, and in 67\% of instances using posterior samples. This demonstrates that posterior disorder samples have greater predictive power than random disorder potentials.

In the real device, we produce two sets of posterior samples from independent iterations of disorder inference. Accepted vectors and the associated labels from both iterations are combined to provide larger sets of results using posterior and random disorders. Examples of 2D current scans which scored highly for double quantum dot features are shown in Fig~\ref{Fig_5}(b). To assess the success of our dot prediction method, a Binomial distribution is fitted to posterior and random results, where the probability of finding a double dot along an accepted voltage vector is $P(\textrm{DD})=p$, with $P(\overline{\text{DD}}) = 1-p$. The fit results in $95\%$ confidence intervals of $0.235\!<\!p_{\textrm{post}}\!<\!0.449$ using posterior samples, and $0.004\! <\!p_{\textrm{rand}}\!<\!0.149$ using random disorders. 

These values, with a separation of the $95\%$ confidence intervals, demonstrate that using posterior disorders results in a higher rate of success than random disorders in finding experimental double quantum dots. Our results show that the inference algorithm produces disorder potentials with predictive power beyond the original domain of training data, and can reduce the human expertise required to tune a double quantum dot.

In addition to the comparison of random and posterior disorders, we also perform the filtering process with featureless (i.e. constant valued) disorder potentials. Fewer vectors are accepted to be tested than when using posterior or random disorders, and vectors which produce the highest scoring 2D scans, identified in results of posterior disorder predictions, are not found. Further details can be found in the supplemental material.

\section{Conclusion}

We demonstrate that hidden disorder in a nanoscale electronic device can be inferred with indirect measurements and physics-aware machine learning. The reparameterisation of the disorder potential proves effective in reducing the dimensionality of the problem, and the successful acceleration of an electrostatic model with a differentiable convolutional neural network allows for Bayesian inference. The entire inference process, from inducing point location optimisation to selecting posterior samples is general and applicable to any gate structure. The device specifics are contained in the electrostatic model, and can easily be adapted to other gate architectures. The prediction of double dot locations using both random and posterior disorders shows the benefits of model assisted tuning, and results indicate that the posterior disorders perform better in this task. The real device still has greater complexity than the model can capture, but the success of current predictions indicates that the use of physics-aware machine learning has narrowed the reality gap. The generality of this method and the minimal data required for inference are promising qualities for future utility in understanding nanoscale quantum devices.

\begin{acknowledgments}
We acknowledge J. Zimmerman and A. C. Gossard for the growth of the AlGaAs/GaAs heterostructure. D.C. would like to thank E. M. Gauger for support and useful discussions on this manuscript. This work was supported by the Royal Society (URF\textbackslash R1\textbackslash 191150), the EPSRC National Quantum Technology Hub in Networked Quantum Information Technology (EP/M013243/1), Quantum Technology Capital (EP/N014995/1), EPSRC Platform Grant (EP/R029229/1), the European Research Council (grant agreement 948932), FQXi Grant Number FQXI-IAF19-01, the Swiss NSF Project 179024, the Swiss Nanoscience Institute, the NCCR SPIN, and the EU H2020 European Microkelvin Platform EMP grant No. 824109. We acknowledge the use of the University of Oxford Advanced Research Computing (ARC) facility in carrying out this work.  
\end{acknowledgments}

\appendix
\renewcommand{\theequation}{\thesection\arabic{equation}}

\section{Self-Consistent Electron Density}
\label{sec:electron_density}

The electron number density is calculated using the Thomas-Fermi approximation in 2D, 
\begin{equation*}
n(\mathbf{r}) = 2\frac{m^*}{\pi\hbar^2}\big(\mu_F-U(\mathbf{r})\big) \Theta\big(\mu_F-U(\mathbf{r}) \big)
\end{equation*}
where $m^*$ is the effective mass of an electron in GaAs, $\mu_F$ is the chemical potential or Fermi level of the 2DEG which is set to zero, and $\Theta(\cdot)$ is the Heaviside step function. The factor of two accounts for spin degeneracy, and the Heaviside step function approximates the Fermi distribution at low temperatures. The electrostatic potential associated with the electron density is 
\begin{equation*}
\phi_{\mathrm{e}}(\mathbf{r}) = -\frac{e}{4\pi\epsilon\epsilon_0}\int\mathbf{dr'} \frac{n(\mathbf{r'})}{\vert\mathbf{r}-\mathbf{r}'\vert}  .
\end{equation*}

A self-consistent solution is computed using an iterative under-relaxation process. The device is fabricated from a wafer (Gossard-060926C) with 2DEG density $n\! =\! 2.64\! \times\! 10^{15}\mathrm{m}^{-2}$ and delta-doping density $n_\mathrm{\delta}\! \approx\! 6\! \times\! 10^{16}\mathrm{m}^{-2}$. As the 2DEG density can be accurately measured, and there is no guarantee that all Si donors become effective dopants, we fit $n_\mathrm{\delta}$ such that the electrostatic model produces the known 2DEG density.  The fitted value is $n_\mathrm{\delta}\! =\! 1.25\! \times\! 10^{16}\mathrm{m}^{-2}$ giving a calculated mean electron density of $\langle n \rangle = (2.64\pm0.04)\!\times\!10^{15}\mathrm{m}^{-2}$, which is the mean and standard deviation uncertainty of 100 calculations. This value is in agreement with the experimental value of $2.64\!\times\!10^{15}\mathrm{m}^{-2}$.

\section{Disorder Covariance}
\label{sec:disorder_covariance}

A Gaussian Process, requiring a covariance function of the random disorder potential, is used to generate random disorder potentials in the inference algorithm.  The donor plane divided into cells, with a random variable $I_{ij}\in\mathbb{N}_0$ determining the number of donors in the cell at $\mathbf{r}_{ij}=(x_i,y_j)$. The potential the 2DEG from the donor ion distribution is given by $\phi_{\mathrm{d}}(\mathbf{r}) = \sum_{ij} I_{ij} \phi_{\mathrm{d}}(\mathbf{r},\mathbf{r}_{ij})$, summing over each cell in the donor plane.

The covariance between two points in the 2DEG plane can be evaluated numerically, and appropriate kernel parameters are fitted. A rational quadratic kernel function
\begin{equation}
\label{ration_quadratic}
k( \mathbf{r},\mathbf{r}' ) = \sigma \bigg(1+\frac{\vert\bm{r}-\bm{r'}\vert^2}{\rho^2}\bigg)^{-1}
\end{equation}
is chosen, with fitted values of $\rho=139.8\mathrm{nm}$ and $\sigma=20.8\mathrm{mV}$. Alternative kernels provide better fits, but the explicit form of the corresponding frequency distribution of random Fourier features is unknown or intractable. 

\section{Reparameterisation}
\label{sec:reparam}

Let $X=\lbrace\mathbf{r}^X_k | k=1,\ldots,n_X\rbrace$ denote the set of dense grid points (34$\times$52 or 45$\times$69 for the experiments in the paper, depending on the CNN model) on the x-y plane, the potential of which is the input of the CNN. Without any measurement, the disorder potential values on $X$, denoted by $\phi^X$, is approximately a random vector following the normal distribution:
\begin{equation*}
\phi^X \sim \mathcal{N}(m\mathbf{1}, K_X),
\end{equation*}
where $m$ is the pre-calculated mean potential level, $\mathbf{1}$ is a one-filled vector, and $K_{X}$ is the covariance matrix, of which element $(i,j)$ is $k(\mathbf{r}^X_i, \mathbf{r}^X_j)$. The value of $m=1184\mathrm{mV}$ is determined from the mean values of 1000 random disorder potentials generated using the electrostatic model (with $\phi_\mathrm{s}$ absorbed into the disorder potential, $m=384\mathrm{mV}$). For the sake of simplicity, the all derivations below are based on the mean-adjusted potential: $\mathbf{f} = \phi^X - m\mathbf{1}$. In order to generate a random sample from $\mathbf{f}$, we can draw a random sample from $\epsilon_X \sim \mathcal{N}(\mathbf{0}, I_X)$ and then transform it as 
\begin{equation}
\mathbf{f} = L_X \epsilon_X,
\label{eqn:prior_exact}
\end{equation}
where $\mathbf{0}$ is a zero-filled vector, $I_X$ is the $n_X \times n_X$ identity matrix, and $L_X$ is the lower Cholesky decomposition of $K_X$.

Since $n_X$ is too large for a practical Bayesian inference problem, and we want to make the inference algorithm independent of $n_X$, the inducing point approach is used. The set of inducing points, $Z=\lbrace \mathbf{r}^Z_k | k=1,\ldots,n_Z \rbrace$, usually has many fewer points than $X$: $n_Z < n_X$. Let $\mathbf{u}$ denote the vector of the mean-adjusted potential values at $Z$ (i.e. $\mathbf{u} = \boldsymbol{\alpha} - m\mathbf{1}$ using notation from the main text). The two mean-adjusted potential vectors $\mathbf{f}$ and $\mathbf{u}$ are jointly a normal distribution, and the joint distribution can be decomposed into two terms: $p(\mathbf{u}, \mathbf{f})=p(\mathbf{u})p(\mathbf{f} | \mathbf{u})$. The first term is a prior distribution, $p(\mathbf{u})=\mathcal{N}(\mathbf{u};\mathbf{0}, K_Z)$, where $K_Z$ is the covariance matrix, of which element $(i,j)$ is $k(\mathbf{r}^Z_i, \mathbf{r}^Z_j)$. The second term is the conditional distribution of $\mathbf{f}$ given $\mathbf{u}$: 
\begin{equation}
p(\mathbf{f} | \mathbf{u}) = \mathcal{N}(\mathbf{f} | K_{XZ} K_Z^{-1}\mathbf{u}, K_X - K_{XZ} K_Z^{-1} K_{ZX}
).
\label{eqn:conditional}
\end{equation}

The computational complexity of computing the covariance of $\mathbf{f} | \mathbf{u}$ is  $O(n_X^3)$ because of the covariance matrix in (\ref{eqn:conditional}). To reduce the computational complexity, any low-rank approximation can be used. In this paper, we approximate the covariance matrix with spectral features. The idea behind this approach is to let inducing points take account of spatially important locations, and the spectral features control relatively unimportant spatial information. The approximation of many types of covariance kernel functions with spectral features is extensively studied in the context of random Fourier features~\cite{Recht2007, Choromanski18a, Sriperumbudur2015, Hensman2018, Avron17a, Li19k}.

The spectral feature is 
$$
\psi(\mathbf{r}) = \frac{1}{\sqrt{q}}[\cos(\omega_1^\top \mathbf{r}), \sin(\omega_1^\top \mathbf{r})\ldots, \cos(\omega_q^\top \mathbf{r}), \sin(\omega_q^\top \mathbf{r})] ,
$$
where $q$ is an arbitrary chosen integer satisfying $n_Z\! <\! 2q\! <\! n_X$, and $w_i$ is a random sample whose probability density function depends on the underlying covariance kernel function. We use $q=300$ in this work. The $n_z\! <\! 2q$ inequality ensures $\Psi_Z\Psi_Z^\top$ (defined below) is invertible, and $2q\! <\! n_x$ ensures that an advantage is gained in computational complexity when using random Fourier features.

The corresponding probability distribution of the samples $\omega_1,\ldots,\omega_q$ given the kernel function (\ref{ration_quadratic}) is
$$
p(\omega) = \rho e^{-\rho \| \omega \|_1},
$$
where $\omega$ is an angular frequency, and $\| \cdot \|_1$ is the L1 norm function.  

The prior covariance matrices, $K_X$ and $K_Z$, are approximated by the spectral features: $K_X\approx \Psi_X \Psi_X^\top$ and $K_Z\approx \Psi_Z \Psi_Z^\top$, where $\Psi_X \in \mathbb{R}^{n_X\times 2q}$ with $\psi(\mathbf{r})^\top$ for $\mathbf{r} \in X$ as rows, and $\Psi_Z$ is defined in a similar fashion. The posterior covariance in (\ref{eqn:conditional}) is approximated as $\mathrm{cov}(\mathbf{f} | \mathbf{u}) \approx \bar{\Psi}_X \bar{\Psi}_X^\top$, where $\bar{\Psi}_X=\Psi_X-\Psi_X\Psi_Z^\top(\Psi_Z \Psi_Z^\top)^{-1}\Psi_Z$. The approximated random field by substituting the approximated covariance matrix into (\ref{eqn:conditional}) is 

\begin{equation}
\mathbf{f}\approx K_{XZ} {L_Z^{-1}}^\top \boldsymbol{\epsilon}_Z + \bar{\Psi}_X \boldsymbol{\epsilon}_{2q} ,
\label{eqn:prior_approx}
\end{equation}
where $\boldsymbol{\epsilon}_Z$ and $\boldsymbol{\epsilon}_{2q}$ are standard normal random vectors with length $n_Z$ and $2q$, respectively, and $L_Z$ is the lower Cholesky decomposition of $K_Z$. The approximated posterior random vector is straightforward,
\begin{equation}
\mathbf{f}|\mathbf{u}\approx K_{XZ} {L_Z^{-1}}^\top \mathbf{u} + \bar{\Psi}_X \boldsymbol{\epsilon}_{2q}.
\label{eqn:post_approx}
\end{equation}
The equation defines the reconstruction of $\phi^X$ through the function
$$
\phi^X \approx f(Z,\mathbf{u}, \boldsymbol{\epsilon}_{2q}) =
K_{XZ} {L_Z^{-1}}^\top \mathbf{u} + \bar{\Psi}_X \boldsymbol{\epsilon}_{2q} + m\mathbf{1},
$$
using the mean adjusted values, and
$$
\hat{\phi} = f(Z,\boldsymbol{\alpha}, \boldsymbol{\beta}) =
K_{XZ} {L_Z^{-1}}^\top (\boldsymbol{\alpha}-m\mathbf{1}) + \bar{\Psi}_X \boldsymbol{\beta} + m\mathbf{1}.
$$
using the notation in (\ref{phi_hat_eqn}) where $\boldsymbol{\alpha} = \mathbf{u} + m\mathbf{1}$, and $\boldsymbol{\beta} = \boldsymbol{\epsilon}_{2q}$.

\section{Detailed Inference Algorithm}
\label{algo_appendix}
\subsection*{Overview}
The posterior inference requires two prerequisites with no interdependence: i) fixing a gate scale factor, ii) optimising the inducing points. The gate voltages are multiplied by the gate scale factor. The gate scale factor is optimised by maximising the likelihood of the observations with assuming the the disorder is perfectly flat. Optimised scale factor values range from 3.48 to 3.94 for runs of the inference algorithm on different thermal cycles.

\subsection*{Inducing Points Optimisation}
Before obtaining any measurements, the inducing point optimisation can be conducted with simulated data. The inducing point optimisation is expensive to compute, but the computation time is not critical, because the optimisation only has to be performed once for a given gate architecture. Algorithm~\ref{alg:IndPtOpt} shows the optimisation procedure. Line number 10 in Algorithm~\ref{alg:IndPtOpt} is important; it generates a posterior random sample with the information $\lbrace Z, \mathbf{u} \rbrace$. The sample only retains information about $\phi_\mathrm{d}$ at $Z$. For the experiments, we used  (\ref{eqn:conditional}) for the posterior sampling, the approximated distribution (\ref{eqn:post_approx}) can be used if the computation speed matters.

For the input of Algorithm~\ref{alg:IndPtOpt}, equation (\ref{eqn:prior_exact}) is used for generating  $\Phi^X_\mathrm{sim}$. Each element of $\mathcal{V}_\mathrm{sim}$ is generated by choosing a disorder randomly from $\Phi^X_\mathrm{sim}$, then choosing a pair of voltage vectors near the closed channel boundary with uniform direction sampling in \cite{moon2020machine}.
The function $\mathrm{KL}(\mathbf{p},\mathbf{p}')$ computes $p_i\log({p_i}/{p_i'}) + (1-p_i)\log({(1-p_i)}/{(1-p_i')})$ element-wise for $\mathbf{p}$ and $\mathbf{p}'$, then it computes the average of them. For the experiments in the paper, $n_d$ and $n_v$ are set to 20, and ADAM optimiser is used. The current probability prediction $\mathcal{F}_U^\mathrm{prob}$ uses the CNN model $\mathcal{F}_U$ and a sigmoid function:
$$
\mathcal{F}_U^\mathrm{prob}(\phi_\mathrm{d}, \mathbf{v})= \sigma_\xi(\mathcal{F}_U(g(\phi_\mathrm{d}, \mathbf{v})); 10),
$$
where $g(\phi_\mathrm{d}, \mathbf{v})$ computes $\phi_\mathrm{in}$ on the dense grid for CNN (see section~\ref{subsec:deep_learning}), and $\sigma_\xi$ is a modified sigmoid function with the steepness parameter, $\sigma_\xi(\cdot; 10)=\xi+ (1-2\xi)\sigma(\cdot; 10)$. The margin $\xi$ allows discrepancy between our approximated model and the real world measurement, and the steepness parameter, set to 10, makes a relatively sharp probability of electric current while allowing the function differentiable. The differentiability is required to use the ADAM optimiser.

\begin{algorithm}[H]
\hspace*{0.02in} {\bf Input:} 
Set of randomly generated disorders $\Phi^X_\mathrm{sim}$, set of randomly generated voltages $\mathcal{V}_\mathrm{sim}$, initial inducing points $Z_\mathrm{init}$, minibatch size of disorders $n_d$, minibatch size of voltages $n_v$, Gaussian process kernel $k$ for disorder, Optimiser parameters $\theta_\mathrm{opt}$, CNN model for current probability prediction $\mathcal{F}_U^\mathrm{prob}$\\
\hspace*{0.02in} {\bf Output: }
Optimised inducing points $Z$
\begin{algorithmic}[1]
\STATE opt $\leftarrow$ Adam optimiser($\theta_\mathrm{opt}$)
\STATE $Z \leftarrow Z_\mathrm{init}$

\WHILE{Stopping criterion not satisfied}
    \STATE $\Phi^X_\mathrm{mini} \leftarrow$ choose random $n_d$ samples from $\Phi^X_\mathrm{sim}$
    \STATE $\mathcal{V}_\mathrm{mini} \leftarrow$ choose random $n_v$ samples from $\mathcal{V}_\mathrm{sim}$
    \STATE loss $\leftarrow$ 0 \;
    \FORALL{$\phi_\mathrm{d}\in\Phi^X_\mathrm{mini}$ and $\mathbf{v}\in\mathcal{V}_\mathrm{mini}$}
        \STATE $\mathbf{p} \leftarrow \mathcal{F}_U^\mathrm{prob}(\phi_\mathrm{d},\mathbf{v})$
        \STATE $\mathbf{u} \leftarrow$ interpolated values of $\phi_\mathrm{d}$ at $Z$
        \STATE $\phi'_\mathrm{d} \leftarrow$ a random sample from the posterior GP with $\lbrace Z,\mathbf{u}  \rbrace$
        \STATE $\mathbf{p'} \leftarrow \mathcal{F}_U^\mathrm{prob}(\phi'_\mathrm{d},\mathbf{v})$
        \STATE $l\leftarrow \mathrm{KL}(\mathbf{p},\mathbf{p}') + \mathrm{KL}(\mathbf{p},\mathbf{p}')$
        \STATE loss $\leftarrow$ loss $+l$
    \ENDFOR
    \STATE $Z \leftarrow$ opt.update(loss, $Z$)
\ENDWHILE

\end{algorithmic}
\caption{Inducing point optimisation}
\label{alg:IndPtOpt}
\end{algorithm}

\clearpage
\subsection*{MCMC Inference}

The goal of the MCMC inference is to generate random samples from the posterior distribution of uncertain variables. The uncertain variables in (\ref{eqn:post_approx}) are $\mathbf{u}$ and $\boldsymbol{\epsilon}_{2q}$. the posterior pdf of $(\mathbf{u},\boldsymbol{\epsilon}_{2q})$ given observed current measurement $D$ is 
\begin{align*}
p(\mathbf{u},\boldsymbol{\epsilon}_{2q} | D)
\propto \prod_{i=1}^n & \mathcal{F}_U^\mathrm{prob}(f(X,\mathbf{u}, \boldsymbol{\epsilon}_{2q}),\mathbf{v}_i)^{y_i} \times \\
& (1-\mathcal{F}_U^\mathrm{prob}(f(X,\mathbf{u}, \boldsymbol{\epsilon}_{2q}),\mathbf{v}_i))^{1-y_i} \times\\
& p(\mathbf{u}) p(\boldsymbol{\epsilon}_{2q}) .
\end{align*}
The prior distributions $p(\mathbf{u})$ and $p(\boldsymbol{\epsilon}_{2q})$ are defined in Appendix \ref{sec:reparam}. For the experiments in the paper, Hamiltonian Monte-Carlo is used with the posterior pdf. Each time MCMC inference is performed a different number of posterior samples are generated. In our work we find typical values to be $150\! <\! n_\mathrm{s}\! <\! 320$.

\bibliographystyle{naturemag}


\newcommand{\beginsupplement}{%
        \setcounter{table}{0}
        \renewcommand{\thetable}{S\arabic{table}}%
        \setcounter{figure}{0}
        \renewcommand{\thefigure}{S\arabic{figure}}%
        \renewcommand{\thesection}{*\Alph{section}}%
       \setcounter{equation}{0}
       \renewcommand{\theequation}{S\arabic{equation}}%
     }

\beginsupplement
\clearpage
\widetext
\begin{center}
\textbf{\large Supplemental Material \\}
\end{center}

\section*{Electrostatic Model}

As discussed in the main text, the total electrostatic potential in the 2DEG $\phi_\mathrm{tot}(\mathbf{r})$ is considered as the sum of components $\phi_\mathrm{tot}(\mathbf{r}) = \phi_{\mathrm{g}}\mathbf{r}) + \phi_{\mathrm{d}}(\mathbf{r}) + \phi_{\mathrm{s}}(\mathbf{r}) + \phi_{\mathrm{e}}(\mathbf{r})$, and the electrostatic potential energy of an electron is $U(\mathbf{r}) = -e\phi_\mathrm{tot}(\mathbf{r})$.

The gate electrodes exist on the surface of the device and the potential from each gate at a depth $d=115\mathrm{nm}$ beneath the gates in the plane of the 2DEG is determined using analytic expressions. A further $5\mathrm{nm}$ is added to the depth of the AlGaAs/GaAs heterojunction ($110\mathrm{nm}$) to account for the extent of the electron density beyond this junction. An representation of the gates is taken form an SEM image a device of identical design. The image of each gate is used and the potential from each pixel is calculated individually and summed to give the total potential from each gate $\phi_{\mathrm{g}_i}(\mathbf{r})$. The total gate potential is
\begin{equation}
\label{gate_potential}
\phi_{\mathrm{g}}(\mathbf{r}) = G_\mathrm{SF} \sum_i v_i \phi_{\mathrm{g}_i}(\mathbf{r}),
\end{equation}

where the sum is over all gates and $v_i$ is the voltage applied to the $i^\text{th}$ gate. The gate scale factor $G_\mathrm{SF}$ is used as the pinned surface model underestimates the magnitude of the gate potential. This underestimation is observed when gate voltages stop current in the experimental device but do not deplete the transport path from source to drain in the simulated device. 

Donor ions exist in a plane at a constant height $h=45\mathrm{nm}$ above the 2DEG. The potential in the 2DEG from a single donor ion at location $\mathbf{r}_{ij}\! =\! (x_i,y_j)$ in the donor plane is
\begin{equation}
\label{single_donor}
\phi_{\mathrm{d}}(\mathbf{r},\mathbf{r}_{ij}) = \frac{e}{4\pi\epsilon\epsilon_0} \Bigg[ \Big(\vert\mathbf{r}-\mathbf{r}_{ij}\vert ^2 + h^2\Big)^{-1/2} -  \Big(\vert\mathbf{r}-\mathbf{r}_{ij}\vert ^2 + (2d-h)^2\Big)^{-1/2}  \Bigg].
\end{equation}

The potential from each donor ion is summed to give the disorder potential $\phi_{\mathrm{d}}(\mathbf{r})$. The electron density is calculated using the Thomas-Fermi approximation in 2D and a self-consistent solution for $\phi_\mathrm{tot}(\mathbf{r})$ is computed using an iterative under-relaxation process.

\section*{Modelling Transport and Dots}

A semi-classical trajectory of electrons between source and drain is calculated by formulating the 2D potential $\phi_\text{tot}$ as a graph $G_{\phi}$, where each pixel is a node. $G_{\phi}$ is defined as
\begin{equation}
\begin{gathered}
G_{\phi} = \big(V,E\big)\ ,\ V = \{v_i\}\ ,\ E = \{e_{ij} \} \ , \\
v_i = U(\mathbf{r}_i)\ ,\ e_{ij} = \frac{1}{2} \big[ v_i + v_j \big] \ ,
\end{gathered}
\end{equation}

where $V$ is the set of nodes in $G_{\phi}$ with $v_i$ the value of the $i^{\mathrm{th}}$ node, and $E$ is the set of edges in $G_{\phi}$ with $e_{ij}$ the edge connecting $v_i$ and nearest neighbour $v_j$.

\begin{figure}[ht]

    \centering
    \includegraphics[width=0.8\textwidth]{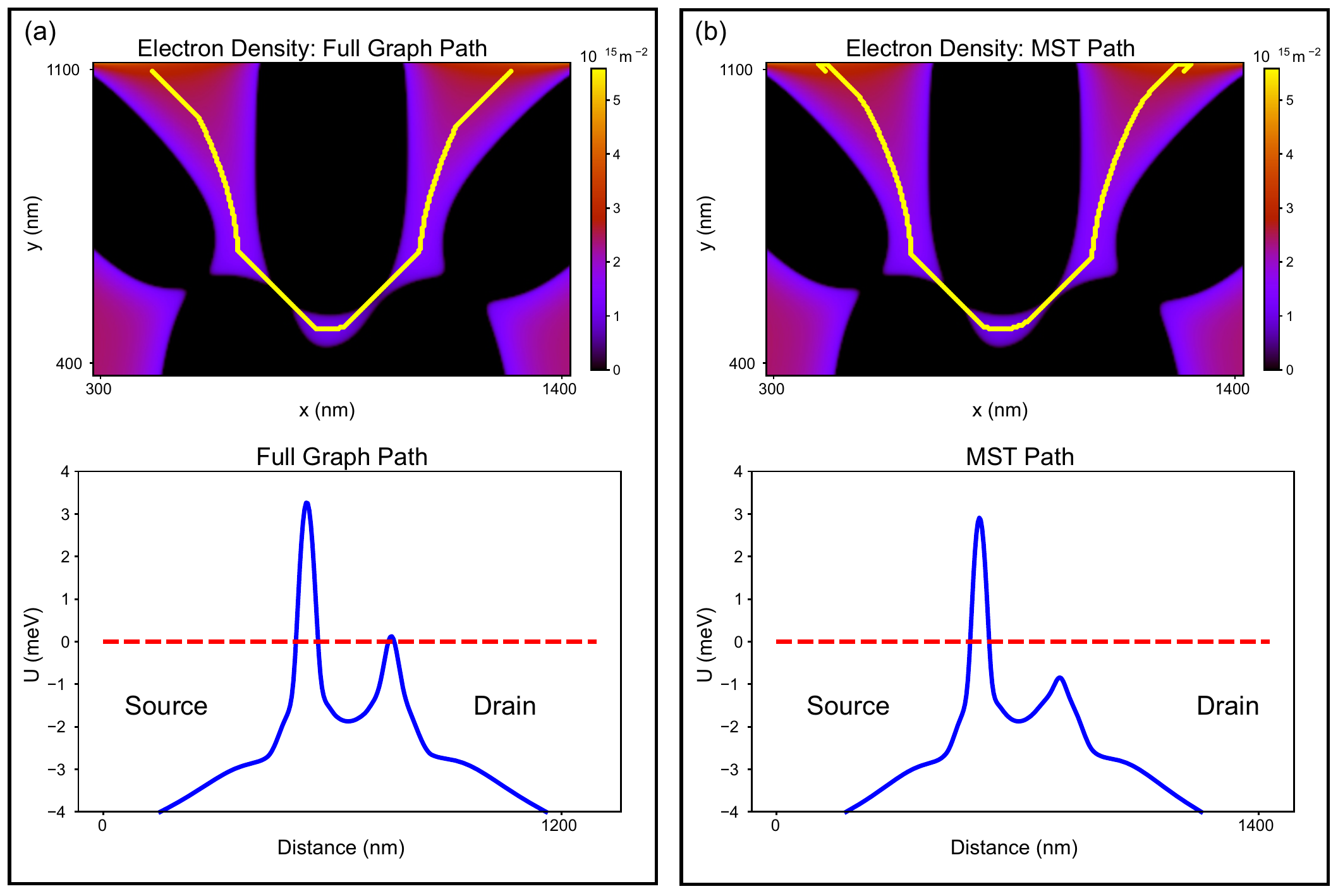}
    
\caption{Comparison of the semi-classical electron trajectory for (a) the full graph $G_{\phi}$ and (b) the MST of $G_{\phi}$ at identical gate voltages. The electron density with the relevant transport path (yellow line) is shown in the upper plot of each panel. The potential along each path (blue line) is shown in the lower plot of each panel, where the Fermi energy, $\mu_F$, is indicated by a red dashed line. The `Distance' axis indicates the total distance moved in 2D space. The transport path using the full graph overestimates the potential barrier height as it shows two regions where $U>\mu_F$. The electron density demonstrates that there is only one region where $U>\mu_F$, matching the prediction of the MST path. }
\label{minimax_path}
\end{figure}

The Dijkstra algorithm results in the shortest path from source to drain through $G_{\phi}$, but this path overestimates the maximum potential energy of an electron along the path. Gate voltages which close the transport channel are then underestimated by our model. This effect can be observed when comparing the transport path with an electron density profile in Figure~\ref{minimax_path}(a). As discussed in the main text, calculating the minimum spanning tree (MST) of $G_{\phi}$ resolves this issue by providing a unique path connecting source and drain with a minimum sum of edge weights. The transport path determined by the MST graph and associated electron density is shown in Figure~\ref{minimax_path}(b). As discussed in the main text, this trajectory allows for the number of quantum dots in the transport channel to be counted.

\clearpage

\section*{Deep Learning}

\begin{figure}[h]

    \centering
    \includegraphics[width=0.5\textwidth]{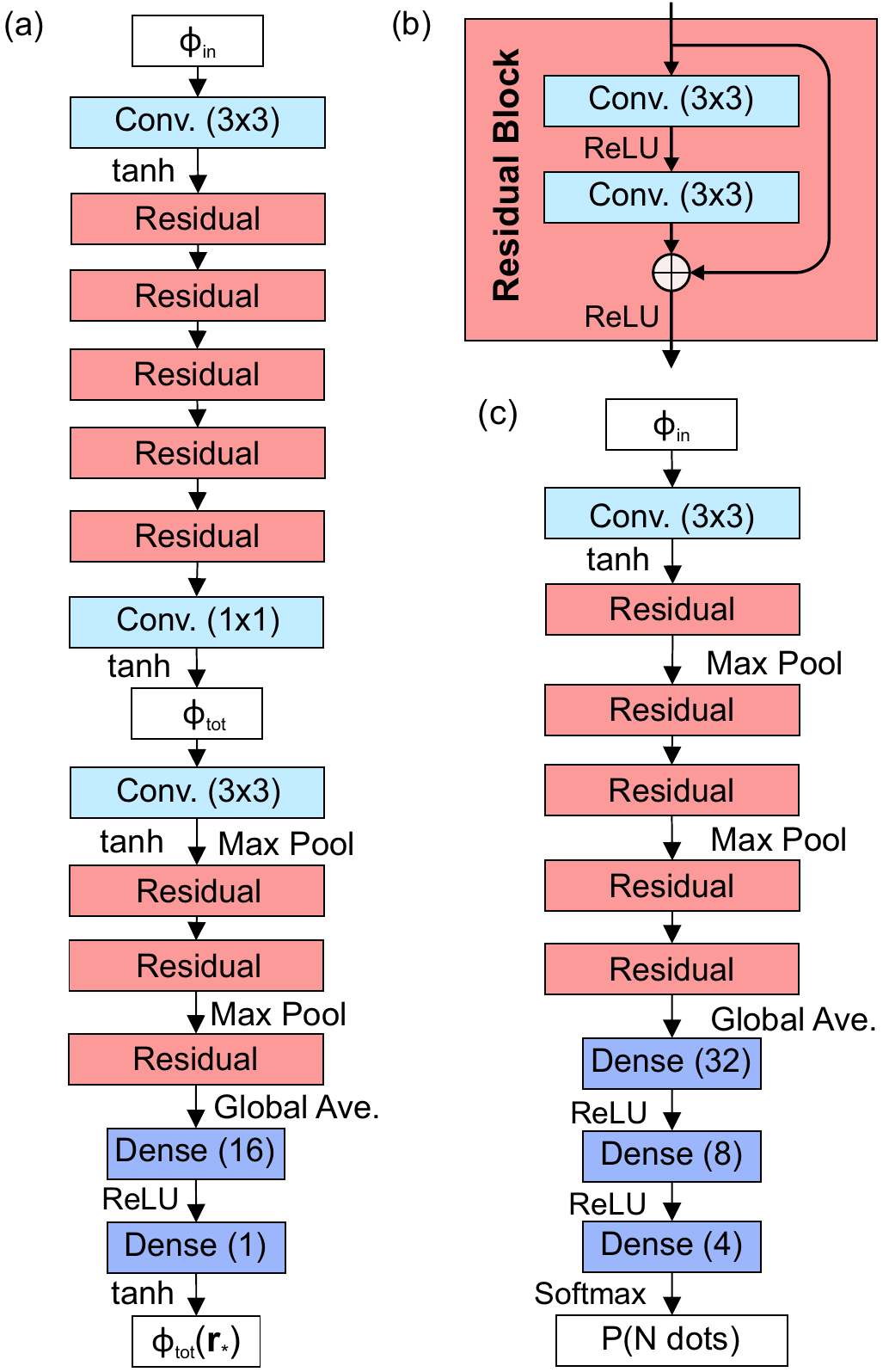}
        
\caption{ The neural networks used to compute $U^*$ and classify dots. Text to the left and right of the arrows indicates the activation function applied and dimension reduction processes respectively. (a) Structure of $\mathcal{F}_U$ beginning with the 2D normalised input potential $\phi_{\mathrm{in}}$, to the self-consistent potential $\phi_{\mathrm{tot}}$, and finally to the minimax value $\phi_{\mathrm{tot}}(\mathbf{r}_*)$. All convolutional units have 32 channels. (b) Schematic of a residual block as used in $\mathcal{F}_U$ and $\mathcal{F}_D$. The number of channels is preserved through a residual block. (c) Structure of $\mathcal{F}_D$ beginning with the normalised input potential, with a probability distribution for the number of dots as the output. All convolutional units have 64 channels, and dots are classified using one-hot encoding for $\{0,1,2,3\}$ dot classes. }
\label{cnn_figure}
\end{figure}

A deep convolutional neural network (CNN), denoted $\mathcal{F}_U$, is trained to approximate $U^*$ given gate voltages and a disorder potential. The structure of $\mathcal{F}_U$ is shown in Figure \ref{cnn_figure}(a). The ResNet skip-connections used in $\mathcal{F}_U$, as shown in Figure \ref{cnn_figure}(b), share similarities with the iterative method used to solve the self-consistent potential.The training data set contains $85,000$ entries with each input being a potential profile $\phi_{\mathrm{in}} = \phi_{\mathrm{g}} + \phi_{\mathrm{d}} + \phi_{\mathrm{s}}$ where $\phi_{\mathrm{g}}$ and $\phi_{\mathrm{d}}$ are randomly generated. The output training data consists of the self-consistent potential $\phi_{\mathrm{tot}}$ and $\phi_{\mathrm{tot}}(\mathbf{r}_*)$ with $U^*=-\phi_{\mathrm{tot}}(\mathbf{r}_*)$. The resolution of each input is reduced from the high resolution required to accurately compute the training data, as shown in Table \ref{network_performance}. Training is performed for 100 epochs with a learning rate of $1\times10^{-3}$ (dropping to $2.5\times10^{-4}$ in two steps) and a MSE loss function. 

The structure of the dot classifier CNN is shown in Figure \ref{cnn_figure}(c). The network learns the mapping $\mathcal{F}_D : \phi_{\mathrm{in}} \to P(N_{\mathrm{dot}})$ for $N_{\mathrm{dot}}\in\{0,1,2,3\}$, where classification is taken as the maximum value of $P(N_{\mathrm{dot}})$. An intermediate classifier is used to generate a suitable dataset, as discussed in the main text. Training is performed for 100 epochs with a learning rate of $1\times10^{-3}$ dropping to $5\times10^{-4}$ after 70 epochs. Test results for dot classification have an accuracy of $98.0\%$ on random data (dot-sparse) and $75.9\%$ on selected (dot-abundant) data.

Following the notation used in Table~\ref{network_performance}, $\mathcal{F}_U$ and $\mathcal{F}_D$ use $\frac{1}{8}$ and $\frac{1}{6}$ resolution of $\phi_{\mathrm{in}}$ respectively. The computation time for both networks $\mathcal{F}_U$ and $\mathcal{F}_D$ is approximately 0.2ms given a 2D potential input and using a GPU. However, processing a vector of gate voltages into a 2D potential increases this time to 0.6ms. The processing involves determining the total gate potential using (\ref{gate_potential}), and summing this with the disorder potential.

\begin{table}[h]
    \renewcommand{\arraystretch}{1.5}
    \centering
    \begin{tabular}{|c|c|c|c|c|}
    \hline
    $\phi_{\mathrm{in}}$ Resolution Fraction                          & $\frac{1}{2}$ & $\frac{1}{4}$ & $\frac{1}{6}$ &  $\frac{1}{8}$ \\
    \hline
    Time per Training Epoch (s)           &  430         &  130          &      69       &   65           \\
    Time per Output (ms)                  &  1.66        &  0.46         &     0.24      &  0.18          \\
    $\phi_{\mathrm{tot}}(\mathbf{r}_*)$ : Test MAE (mV)                         &  1.41        &  1.18         &     1.25      &  1.27          \\
    Transport Channel : Test Error ($\%$)             &  1.57        &  1.23         &     0.96      &  1.20          \\
    
    \hline
    \end{tabular}
    \caption{Performance metrics of $\mathcal{F}_U$ for different  resolutions of $\phi_{\mathrm{in}}$, computed using a GPU (GTX 1080 Ti). Max-pooling processes are adapted for each resolution, otherwise the networks are identical. The resolution reduction fraction is applied to both input dimensions. Full resolution $(269,411)$ is only required for computing training data, so not considered. Time per output is from a batch of 1000 inputs. MAE is chosen so that units remain in mV, and transport channel error is based on binary classification using the value of $\phi_{\mathrm{tot}}(\mathbf{r}_*)$ produced by $\mathcal{F}_U$. }
    \label{network_performance}
    \renewcommand{\arraystretch}{1}
\end{table}

\clearpage
\section*{Disorder Covariance}

As discussed in the main text, the potential at a point $\mathbf{r}$ in the 2DEG from the donor ion distribution is given by $\phi_{\mathrm{d}}(\mathbf{r}) = \sum_{ij} I_{ij} \phi_{\mathrm{d}}(\mathbf{r},\mathbf{r}_{ij})$, summing over each cell in the donor plane (subscripts $i$ and $j$ reflect the 2D grid of cells). The covariance of $\phi_{\mathrm{d}}$ between two points in the 2DEG plane can be computed as 
$
\mathrm{cov}\big(\mathbf{r},\mathbf{r}'\big) = \mathrm{var}(I)\sum_{ij}\phi_{\mathrm{d}}(\mathbf{r},\mathbf{r}_{ij})\phi_{\mathrm{d}}(\mathbf{r}',\mathbf{r}_{ij})
$, where $I$ is the distribution from which each $I_{ij}$ is independently drawn. Using the correlation function eliminates the dependence on $I$,
\begin{equation}
\label{correlation_function}
\mathrm{corr}\big(\mathbf{r},\mathbf{r}'\big) = \frac{\sum_{ij}\phi_{\mathrm{d}}(\mathbf{r},\mathbf{r}_{ij})\phi_{\mathrm{d}}(\mathbf{r}',\mathbf{r}_{ij})} {\sqrt{\sum_{ij}\phi_{\mathrm{d}}(\mathbf{r},\mathbf{r}_{ij})^2} \sqrt{\sum_{ij}\phi_{\mathrm{d}}(\mathbf{r}',\mathbf{r}_{ij})^2}}.
\end{equation}

This expression is evaluated numerically, and appropriate parameters are fitted to the kernel function given in the main text, $k( \mathbf{r},\mathbf{r}' ) = \sigma \big(1+\frac{\vert\bm{r}-\bm{r'}\vert^2}{\rho^2}\big)^{-1}$. Normalising to the correlation kernel, a least squares fit results in $\rho=139.8\mathrm{nm}$, with a MSE of $1.65\times10^{-4}$. An appropriate value of $\sigma=20.8\mathrm{mV}$ is found using the standard deviations of 1000 disorder profiles.


\section*{Inducing Point Values}

\begin{figure}[h]

    \centering
    \includegraphics[width=0.9\textwidth]{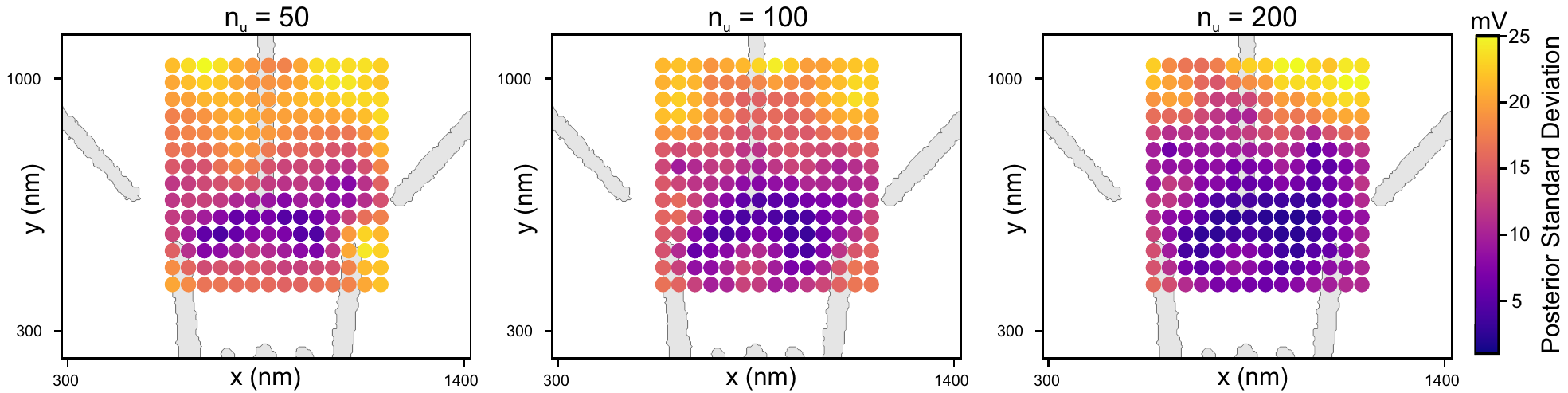}
        
\caption{ Disorder inference results using simulated data for three training dataset sizes ($n_\mathrm{u}=50,100,200$), with $D=\lbrace(\mathbf{v}_i, y_i) | i=1,\ldots,2n_\mathrm{u} \rbrace$ as discussed in the main text). The inducing point locations are indicated by circles with the gate structure in the background. The colour of each inducing point represents the standard deviation of the posterior disorder potentials at that point. }
\label{posterior_std}
\end{figure}

As discussed in the main text, the standard deviation of inducing point values across the posterior samples indicates how much the inference algorithm has learned about the disorder potential at those locations. A low standard deviation on an inducing point means that all posterior samples have similar values at that location and therefore the inference algorithm is confident of the disorder potential there. To further demonstrate this, we consider inference results for the simulated device using training datasets of different sizes as shown in Figure~\ref{posterior_std}. The training dataset is $D=\lbrace(\mathbf{v}_i, y_i) | i=1,\ldots,2n_\mathrm{u} \rbrace$, as defined in the main text. A training dataset with $n_\mathrm{u}=200$ is used for inference results discussed in the main text.

We observe that the number of inducing points with a low posterior standard deviation increases with the size of the training dataset. This indicates that the inference algorithm has gained information about a larger area of the disorder potential by considering more directions in voltage space. We can also observe that even for a small training dataset, the inference results are most confident about the disorder potential values at the tip of gate G1 which reflects its role in depleting the electron density along the path from source to drain.

The lowest standard deviation on a given inducing point is approximately 2mV using the simulated measurements to inform the inference algorithm, and approximately 5.5mV using experimental measurements. This performance difference can be expected as the simulated device is a controlled and self-contained environment, whereas the experiment will have more unknowns which may not be accounted for in the model.  

\clearpage
\section*{Posterior Disorder Samples}

\begin{figure}[h]

    \centering
    \includegraphics[width=0.98\textwidth]{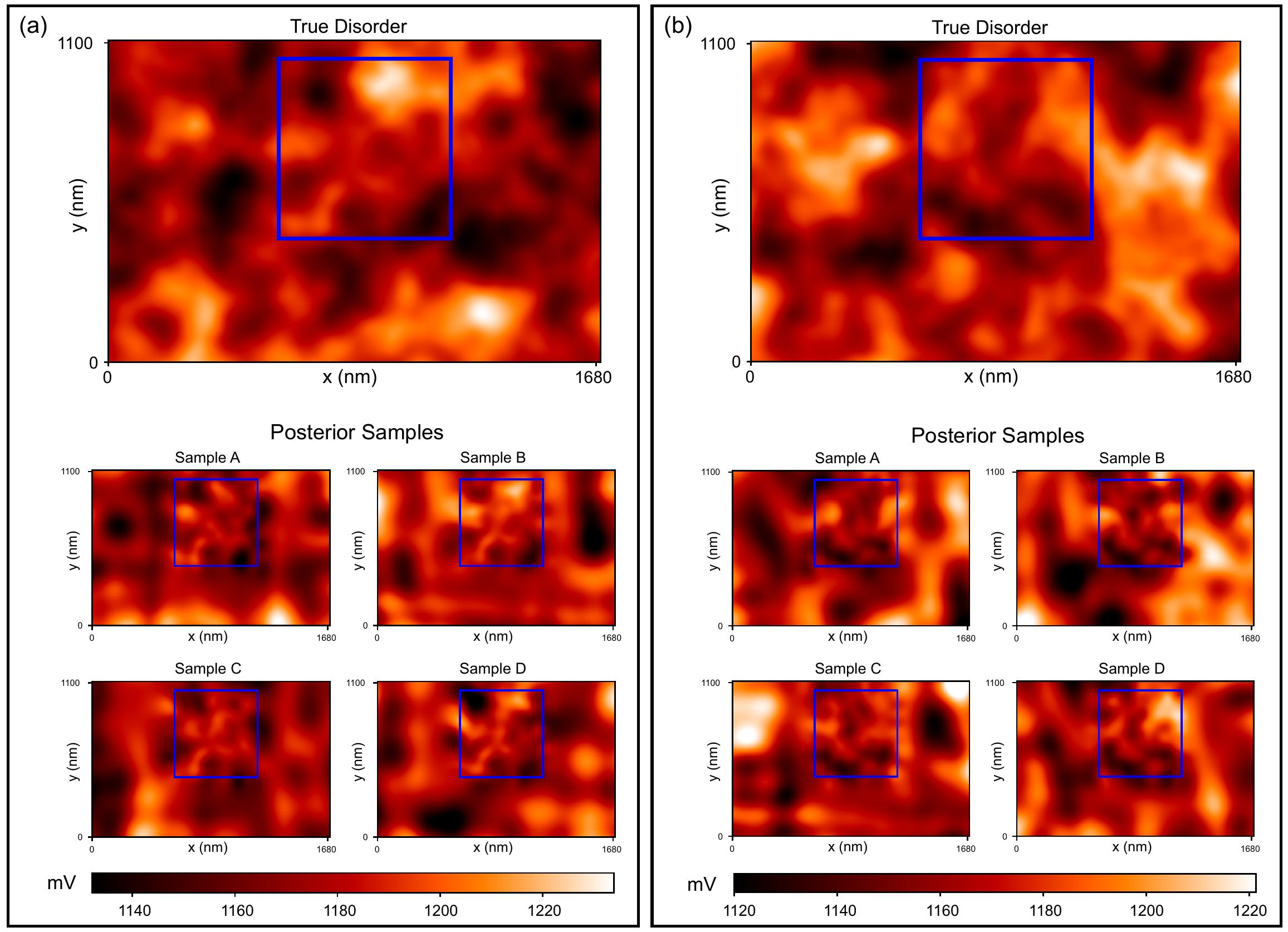}
        
\caption{ The true disorder and 4 randomly selected posterior disorder samples for 2 iterations of the inference algorithm on a simulated device, (a) and (b). The true disorder is randomly generated using the electrostatic model with randomly located donor ions. The blue box on each disorder potential indicates the region spanned by the optimised inducing points. The posterior disorder samples are the same resolution as the true disorder, (134x206). In each subfigure, the posterior disorders have more detailed features within the box, and qualitative similarities with the true disorder can be observed. All plots in each subfigure share the same scale, as shown at the bottom of each subfigure. }
\label{posterior_disorders}
\end{figure}

The inference algorithm produces a set of posterior values for the inducing point values, $\boldsymbol{\alpha}$, and random Fourier feature amplitudes, $\boldsymbol{\beta}$. These values are used to produce posterior samples of the real-space disorder potential as outlined in the main text and Appendix~\ref{sec:reparam}. The resolution of these posterior disorder potentials can be chosen depending on the desired use (e.g. as inputs to $\mathcal{F}_\mathrm{U}$, $\mathcal{F}_\mathrm{D}$, or the self-consistent electrostatic model). Figure~\ref{posterior_disorders} shows the true disorder and posterior samples for two iterations of the inference algorithm on a simulated device. The posterior samples exhibit much more detailed features inside the region spanned by the optimised inducing points where qualitative similarities with the true disorder observed. This further demonstrates the information gained at these points (in addition to Figure~\ref{posterior_std}).

For posterior disorder potentials, features outside the inducing point region are governed by the amplitudes of random Fourier features contained in $\boldsymbol{\beta}$ which are necessary to ensure the posterior samples are continuous and suitable to be used as inputs to $\mathcal{F}_\mathrm{U}$ and $\mathcal{F}_\mathrm{D}$.

\clearpage
\section*{Double Dot Filtering}

As discussed in the main text, the posterior disorder samples can be used to predict quantum dot locations by filtering random unit vectors. Filtering involves scanning along a random voltage vector from the origin and using the maximum $P(N_{\mathrm{dot}})$ computed by $\mathcal{F}_D$ along that vector for each disorder sample. If $\max{P(N_{\mathrm{dot}})}>0.5$ along a vector for a given disorder sample, the score associated with the vector is increased by one. If the score is greater than a given threshold after using all the disorder samples, the vector is accepted and taken to a test device. 

We perform 5 independent iterations of the inference algorithm on a simulated device, where each iteration uses a different true disorder potential. The output of $\mathcal{F}_D$ with the true disorder is used as the ground truth when determining whether a vector produces a double dot. For each true disorder potential we generate a dataset of random unit vectors where exactly half of the vectors show double dots (dataset size is either 500 or 1000 vectors depending on the iteration). This distribution of vectors producing double dots is selected to give a better representation of classification statistics when performing vector filtering. Without this selection, only $0.83\%$ of randomly generated unit vectors produce a double quantum dot. For each set of random unit vectors, we perform filtering using the corresponding set of $n_\mathrm{s}$ posterior disorder samples, $n_\mathrm{s}$ random disorders, and $n_\mathrm{s}$ featureless disorders. The value of $n_\mathrm{s}$ changes for each iteration. Using the simulated device, this filtering process allows us to determine the mean false positive rate (FPR) and mean true positive rate (TPR) across the 5 iterations when using different thresholds to determine whether a vector is accepted. 

\begin{figure}[h]

    \centering
    \includegraphics[width=0.98\textwidth]{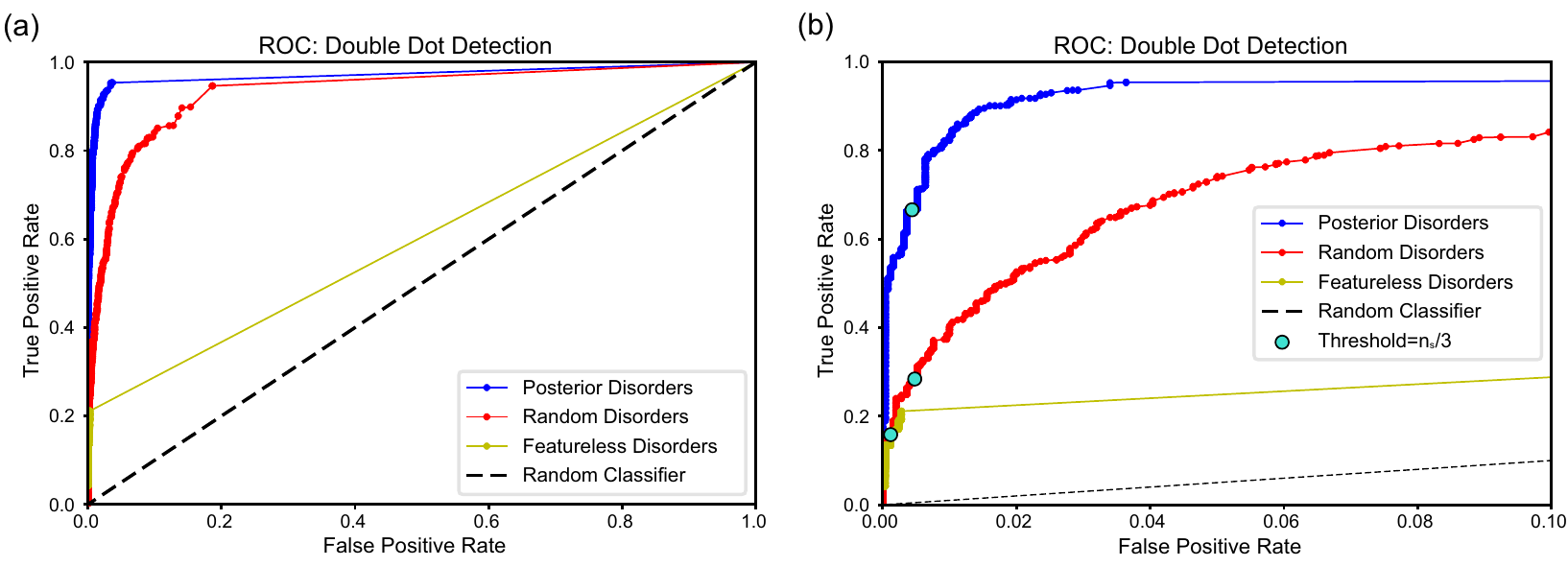}
        
\caption{ The receiver operator characteristic (ROC) curve, comparing the false positive rate (FPR) and true positive rate (TPR) across the domain of possible thresholds $[0,n_\mathrm{s}]$ for $n_\mathrm{s}$ disorder samples. The mean FPR and TPR from 5 independent iterations of the inference algorithm and filtering process is shown. \textbf{a} The ROC curves for the double dot filter process using posterior (blue line), random (red line), and flat (yellow line) disorders, along with the ROC curve associated with a random classifier (dashed line). \textbf{b} The same ROC curves as shown in \textbf{a}, but with a truncated false positive axis. The location corresponding to a threshold of $n_\mathrm{s}/3$ as used in the experiment is indicated by a turquoise circle on each curve.  }
\label{ROC_curve}
\end{figure}

We use the receiver operator characteristic (ROC) curve for posterior, random, and featureless disorders to identify suitable values for the threshold, as shown in Figure~\ref{ROC_curve}(a). A low FPR and high TPR are features of a good classifier. Posterior disorders and random disorders both perform much better than featureless disorders and a random classifier. Posteriors perform better than random disorders. This is further evidence of the success of our inference algorithm. A more detailed comparison in Figure~\ref{ROC_curve}(b) shows that the posterior curve has a much sharper increase in true positive rate. We choose a threshold value of $n_\mathrm{s}/3$ for the experiment as it allows both posteriors and random disorders to have a moderate true positive rate, while maintaining a low false positive rate. A low false positive rate is desirable because experimental 2D current scans required for human labelling are slow. Other threshold values can be chosen depending on the desired acceptance rate.

For the real device, as discussed in the main text, we generate 5000 random unit vectors which are filtered using posterior and random disorders. The mean number of accepted unit vectors across 2 iterations of filtering is 38 using posterior disorders, and 26 using random disorders. Featureless disorder potentials are also used to filter vectors for testing on the real device. We use the optimised scale factor from each iteration of the inference algorithm along with 100 featureless disorders, where the constant value is determined by the mean of a random disorder potential. The mean number of accepted unit vectors across the 2 iterations is 8.5 using featureless disorders, indicating that double dots are not often formed in the model using only the gate potentials. Of the 11 unique vectors which are accepted by featureless disorders across both iterations of filtering, 2 are labelled as producing quantum dots by human experts. Filtering using featureless disorders fails to find the top scoring vectors for double quantum dots, as identified by the posterior samples. One of the accepted vectors was not tested due to experimental difficulties. 

\clearpage
\section*{Current Scans}

\begin{figure}[h]

    \centering
    \includegraphics[width=0.70\textwidth]{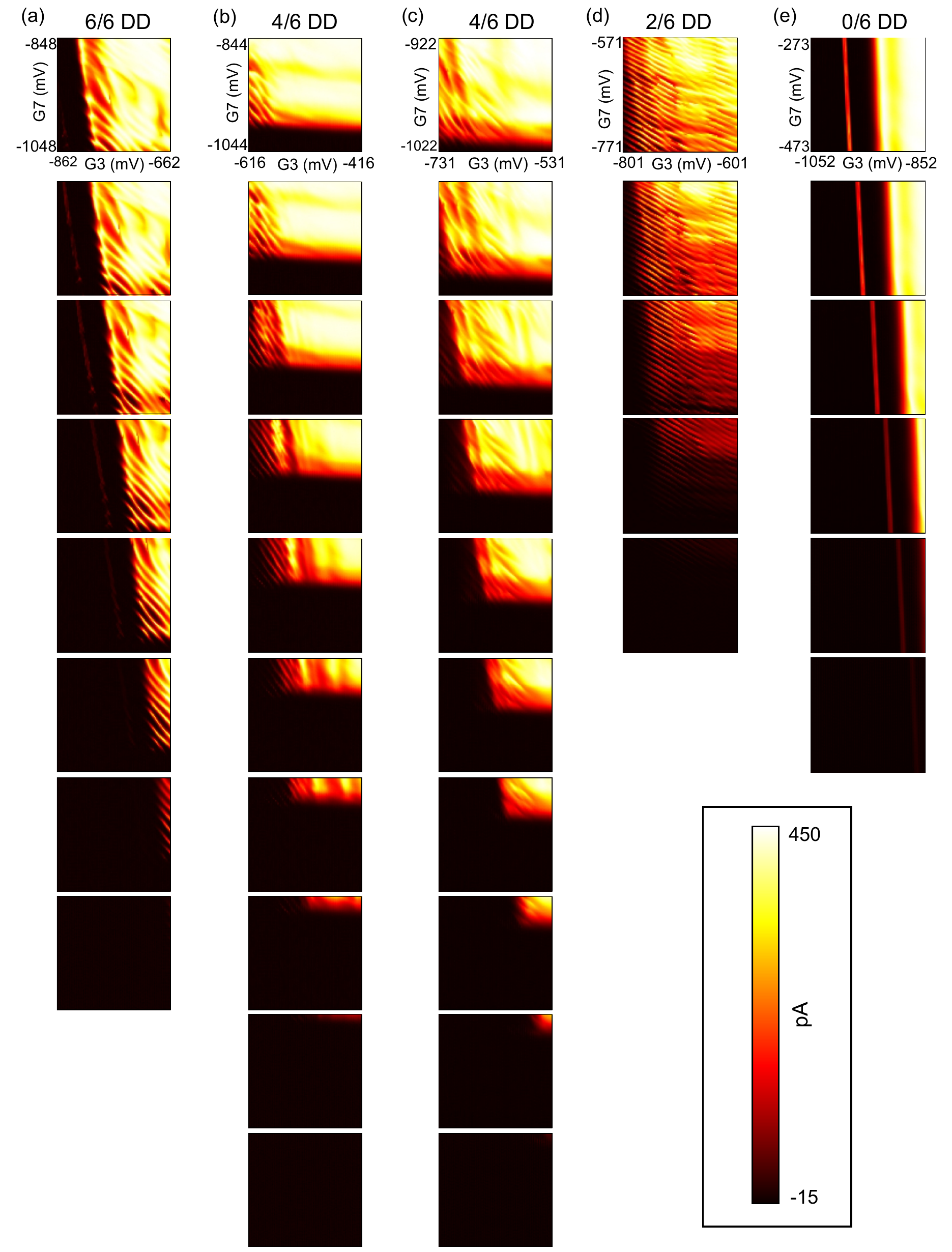}
        
\caption{ Each column \textbf{a}-\textbf{e} is a set of 2D current scans taken using a vector accepted by the filtering process discussed in the main text. Each scan is a 200mV$\times$200mV window over gates G3 and G7, with the original 7-dimensional voltage vector at the centre. The magnitude of R (as defined in the main text) is increased by 13.3mV from one row to the next. The axes of the first scan in each column are labelled to indicate the range of voltage values considered. The colour scale shared by all plots is shown in the box. The result of human expert labels is indicated at the top of each column. For example, `4/6 DD' indicates that 4 out of 6 labellers considered a double dot to exist in the set of current scans.  }
\label{current_scans}
\end{figure}

When scanning an accepted voltage vector in the experiment, 2D current scans are taken at  13.3mV intervals in $R$ once the current drops to 80\% of the open channel value. The protocol stops taking scans once the maximum current value in a 2D scan is less than 100pA. The set of 2D scans corresponding to each vector is labelled by 6 human experts to determine whether a double dot exists along that vector. If more than 3 out of the 6 human experts label a double dot in a set of scans, then a double dot is deemed to exist in transport measurements along the given vector. In the event of a split decision (i.e. 3 out of 6 double dot labels), we do not consider a double dot to exist. Examples of these 2D scans and the number of double quantum dot labels from the human experts are shown in Figure~\ref{current_scans}. 

\clearpage

\end{document}